\documentclass[prd,aps,twocolumn,a4paper,showkeys,nofootinbib,10pt]{revtex4-1}

\usepackage{amsmath}
\usepackage{amsfonts}
\usepackage{amssymb}	
\usepackage{graphicx}
\usepackage{bm}
\usepackage{color}
\usepackage{commath}
\usepackage{hyperref}
\usepackage[T1]{fontenc}
\usepackage{xcolor}
\usepackage{verbatim}
\usepackage{comment}
\usepackage{float}
\usepackage{subfig}
\usepackage{amsbsy}
\usepackage{makecell}
\usepackage{multirow}

\usepackage[normalem]{ulem}
\allowdisplaybreaks

\newcommand{\orcid}[1]{\href{https://orcid.org/#1}{
\includegraphics[width=10pt]{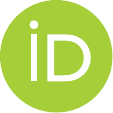}
}}

\begin{document}
\title{Charged black-hole binary radiation at second post-Newtonian order}

\author{Andrea Placidi${}^{1}$\orcid{0000-0001-8032-4416}}
\author{Elisa Grilli${}^{1,2}$\orcid{0009-0007-7529-4962}}
\author{Matteo Pegorin${}^{3,4,5}$\orcid{0009-0003-1248-871X}}
\author{Marta Orselli${}^{1,2}$\orcid{0000-0003-3563-8576}}


\affiliation{${}^1$Dipartimento di Fisica e Geologia, Universit\`a di Perugia, I.N.F.N. Sezione di Perugia, \\ Via Pascoli, I-06123 Perugia, Italy}
 \affiliation{${}^2$ Center of Gravity, Niels Bohr Institute, Copenhagen University,\\ Blegdamsvej 17, DK-2100 Copenhagen \O{}, Denmark}
\affiliation{${}^3$ Dipartimento di Fisica e Astronomia ``Galileo Galilei'', Universit{\`a} degli Studi di Padova, via Marzolo 8, I-35131 Padova, Italy}
\affiliation{${}^4$INFN, Sezione di Padova, Via Marzolo 8, I-35131 Padova, Italy}
\affiliation{${}^5$Max Planck Institute for Gravitational Physics (Albert Einstein Institute), Am M{\"u}hlenberg 1, Potsdam 14476, Germany}

\email{andrea.placidi@unipg.it}
\email{elisa.grilli@nbi.ku.dk}
\email{matteo.pegorin.1@phd.unipd.it}
\email{orselli@nbi.ku.dk}

\begin{abstract}
In this work, we build on Ref.~[Phys.~Rev.~D 112, 124060 (2025)], where we derived the dynamics of an electrically charged binary system at second post-Newtonian (2PN) order, by extending the analysis, at the same PN accuracy, to the associated emission of electromagnetic and gravitational radiation. We focus on non-spinning binaries that evolve quasi-adiabatically along a sequence of circular orbits, and compute the 2PN expression for the corresponding total energy flux at infinity. With respect to the well-known case of a neutral binary, the presence of charge induces both a new leading-order electromagnetic contribution of dipolar nature and additional charge-dependent corrections in the gravitational sector. To account for this richer phenomenology, we extend both the PN-matched multipolar post-Minkowskian and the effective field theory approaches to the radiation sector of charged binaries, with exact agreement for the contributions computed independently in both formalisms.
We derive the radiative symmetric-trace-free multipole moments of the vector and gravitational fields at 2PN order.
In doing so, we also compute 2PN-accurate charge-dependent corrections to the spherical multipoles of the waveform emitted by the binary, including a new memory contribution sourced by the stress-energy tensor of the electromagnetic radiation. These results lay the foundation for developing accurate waveform models capable of assessing the observational signatures of electric charge in gravitational-wave signals.


\end{abstract}

\maketitle
\newpage
\section{Introduction}
\label{sec:intro}

The direct observation of gravitational waves (GWs) from compact-binary coalescences has opened a precision era in strong-field gravity. The interpretation of these signals relies on accurate waveform models, which in turn require a systematic description of both the conservative dynamics of the source and the radiation emitted at infinity. In general relativity (GR), this program has been developed to high post-Newtonian (PN) accuracy for neutral compact binaries, using a combination of near-zone PN methods, multipolar post-Minkowskian (MPM) wave generation, numerical relativity, and effective field theory techniques~\cite{Blanchet:2013haa,Berti:2015itd,Will:2014kxa, Goldberger:2004jt, Futamase:2007zz,Blanchet:2013haa,Porto:2016pyg,Schafer:2018kuf,Levi:2018nxp,Jaranowski:1997ky,Damour:2014jta,Jaranowski:2015lha,Bernard:2015njp,Bernard:2016wrg,Damour:2016abl, Blanchet:2023sbv,Blanchet:2023bwj,
Goldberger:2004jt,Kol:2013ega,Foffa:2012rn,Foffa:2019rdf,Foffa:2016rgu,Foffa:2019yfl,Blumlein:2020pog,Blumlein:2020znm,
Foffa:2019hrb,Blumlein:2019zku,Foffa:2020nqe,Blumlein:2020pyo,Blumlein:2021txe,Porto:2024cwd,
Blumlein:2021txj,
Porto:2005ac,Levi:2015msa,Kim:2021rfj,Levi:2020uwu,Levi:2019kgk,Kim:2022pou,Kim:2022bwv,Levi:2022dqm,Levi:2022rrq,Mandal:2022nty,Mandal:2022ufb,Mandal:2023lgy,Mandal:2023hqa,
Goldberger:2009qd,Ross:2012fc,
Cho:2021mqw,Cho:2022syn,Amalberti:2023ohj,Amalberti:2024jaa,Mandal:2024iug, Brunello:2025gpf,Placidi:2025lxn,Brunello:2026anu,Bini:2026vaq, DiRusso:2026fqn, Bini:2026dvn, Bini:2026oie, Bini:2026suo,Almeida:2026clf,Porto:2026fsd}, as well as in the post-Minkowskian regime~\cite{Westpfahl:1979gu,Westpfahl:1980mk,Bel:1981be,Westpfahl:1985tsl,schafer1986adm,Ledvinka:2008tk,Buonanno:2022pgc,Travaglini:2022uwo,Bjerrum-Bohr:2022blt,Kosower:2022yvp,Damour:2016gwp,
Cheung:2018wkq,Bern:2019nnu,Bern:2022kto,Bern:2024adl,Kosower:2018adc,Bjerrum-Bohr:2018xdl,Cristofoli:2019neg,Damgaard:2019lfh,Brandhuber:2021eyq,Vines:2017hyw,Kalin:2020mvi,Kalin:2020fhe,Mogull:2020sak,Jakobsen:2021zvh,Jakobsen:2022psy,Driesse:2024xad,Driesse:2024feo,Bern:2025zno,Brunello:2026rdk,Brunello:2026lzf, DeAngelis:2026ymn, Bohnenblust:2026ujk, Jakobsen:2026ayw, Driesse:2026qiz, Bern:2025wyd,Dlapa:2026oyq,Correia:2024jgr,Caron-Huot:2025tlq,Correia:2026utp}.

In standard astrophysical environments, black holes are expected to be electrically neutral to a very good approximation. Even when external electromagnetic (EM) fields are present, charge separation is limited by selective accretion, plasma effects, vacuum breakdown, and pair creation, which tend to discharge an astrophysical black hole on short timescales~\cite{Gibbons:1975kk,Wald:1974np,Blandford:1977ds,Palenzuela:2010nf}. This expectation is built into the standard modeling of binary black hole signals by the LIGO--Virgo--KAGRA network~\cite{LIGOScientific:2014pky,Virgo:2014yos,LIGOScientific:2021djp}. Nevertheless, charged black holes provide a useful and well-controlled arena for testing the Kerr-Newman sector of Einstein-Maxwell theory and for parameterizing possible beyond-GR or hidden-sector effects. In particular, the effective ``charge'' entering the spacetime may represent magnetic monopole charge, a charge under a dark or hidden $U(1)$ interaction, or a vector charge in modified-gravity scenarios~\cite{Preskill:1984gd,Cardoso:2016olt,Bozzola:2020mjw,Bozzola:2021elc}. In such cases, the usual EM discharge arguments need not apply, and gravitational-wave observations can be used to constrain the corresponding charge-to-mass ratios~\cite{Abbott:2016blz,Yunes:2016jcc,Carullo:2021dui}.

Charged compact binaries exhibit a richer structure than their neutral counterparts. At the conservative level, the Coulomb interaction modifies the binary dynamics already at Newtonian order. At the dissipative level, the vector field gives rise to dipolar radiation, which enters one PN order earlier than the leading quadrupolar gravitational-wave flux. Moreover, the electromagnetic field also contributes to the stress-energy tensor that sources the metric, leading to charge-dependent corrections in the gravitational radiation itself. Several aspects of this problem have been studied in the PN and effective-one-body frameworks, including Einstein-Maxwell-dilaton binaries, charged-binary dynamics, and wave generation in Einstein-Maxwell theory~\cite{Julie:2017rpw,Khalil:2018aaj,Patil:2020dme,Gupta:2022spq,Henry:2023guc,Henry:2023len, Wilson-Gerow:2023syq,Alonzo-Artiles:2026wbe}.

This work is the second part of a program aimed at developing a 2PN-accurate description of electrically charged, non-spinning black-hole binaries. In Ref.~\cite{Placidi:2025xyi}, we derived the conservative dynamics at 2PN order, including the equations of motion, the transformation to the center-of-mass (CoM) frame, and several gauge-invariant observables. We also computed the leading dissipative effects associated with EM dipole radiation. Here we complete the analysis by deriving the radiation field at infinity, the waveform modes, and the corresponding total energy flux for quasi-circular inspirals.

Our calculation combines two complementary formulations of the PN expansion.
We extend the PN-matched MPM formalism to Einstein-Maxwell theory, and use it to relate the near-zone source multipole moments to the radiative multipole moments measured at future null infinity, while systematically accounting for nonlinear propagation effects such as tails and memory~\cite{Blanchet:1987wq,Blanchet:1992br,Blanchet:1998in,Blanchet:2013haa,Henry:2023len}. We first compute the electric and magnetic source moments of the vector field and construct the corresponding radiative moments, including the leading vector-field tail. We then compute the charge-dependent corrections to the gravitational mass- and current-type source moments. These corrections arise both directly from the EM components of the effective stress-energy densities and indirectly through the charge-dependent order reduction and CoM transformation. We also derive the associated radiative gravitational moments, including the usual tail terms and new nonlinear contributions sourced by the electromagnetic stress-energy tensor.
We also extend the effective field theory approach~\cite{Goldberger:2004jt,Goldberger:2009qd,Ross:2012fc,Leibovich:2019cxo,Amalberti:2023ohj,Amalberti:2024jaa,Mandal:2024iug} to charged systems. This is used to independently compute the subset of source multipole required to compute the instantaneous energy flux through relative 2PN order with respect to the leading electromagnetic dipole flux.

Specializing to circular orbits, we obtain the spin-weighted spherical-harmonic modes of the gravitational waveform and the total energy flux through 2PN accuracy. The total flux naturally splits into a vector contribution, dominated by the leading electric dipole, and a tensor contribution, which reduces to the standard gravitational-wave flux in the neutral limit. We verify that our results reproduce the known 1PN charged-binary fluxes~\cite{Khalil:2018aaj,Julie:2018lfp} and the standard neutral-binary expressions~\cite{Blanchet:1995fg,Blanchet:2001aw}. Finally, we illustrate the impact of the charge-dependent corrections on the flux for representative charge configurations, including cases in which one of the black holes approaches extremality.

The paper is organized as follows. In Sec.~\ref{sec: setup}, we summarize the Einstein-Maxwell setup, introduce the PN-expanded field equations, and derive the effective source functions needed for the computation. We also recall the circular-orbit reduction and the gauge-invariant frequency parameter used throughout the paper. In Sec.~\ref{sec:ElectricMagneticMultipoles}, we compute the electric and magnetic multipole moments of the vector field and discuss their nonlinear tail corrections. In Sec.~\ref{sec:MassCurrentMultipoles}, we derive the charge-dependent gravitational source moments and construct the corresponding radiative moments, including tail and EM-memory effects. In Sec.~\ref{sec:waveformModes}, we project the radiative symmetric-tracefree (STF) moments onto spin-weighted spherical harmonics and obtain the waveform modes. In Sec.~\ref{sec:EnergyFlux}, we compute the vector and tensor contributions to the energy flux and study their behavior for selected charge configurations. We conclude in Sec.~\ref{sec:conclusion}. Lengthy expressions are collected in the appendices. The explicit results for the electric and magnetic source moments, the mass and current source moments, the GW modes, and the vector and tensor energy flux are also provided in electronic form in the Supplementary Material~\cite{SupplementalMaterial} accompanying this paper.

\subsection{Notation and conventions}
\label{subsec: notation}

The subject of our analysis is a binary system composed of two black holes with masses $m_A$ and charges $q_A$, with $A=1,2$. For later convenience, the two black holes are labeled so that $m_1 \ge m_2$.

Adopting harmonic coordinates, we denote by $\mathbf{y}_A(t)$ their position at time $t$, $\mathbf{v}_A= d\mathbf{y}_A(t)/dt$ their coordinate velocities, and $\mathbf{a}_A= d\mathbf{v}_A(t)/dt$ their coordinate accelerations. The harmonic-coordinate distance between the two bodies is defined by $r=|\mathbf{y}_1(t)-\mathbf{y}_2(t)|$.
More in general, the relative binary's separation, velocity and acceleration read
\begin{align}
    &\mathbf{r}=\mathbf{y}_1-\mathbf{y}_2 \quad \text{with\, } r=|\mathbf{r}|\text{\, and\, } \mathbf{n}=\frac{\mathbf{r}}{r} \\ 
    \mathbf{v}&=\frac{d \mathbf{r}}{dt}= \mathbf{v}_1-\mathbf{v}_2, \qquad\mathbf{a}=\frac{d \mathbf{v}}{dt}= \mathbf{a}_1-\mathbf{a}_2.
\end{align}
The position of the field point in harmonic coordinates is instead denoted by $\mathbf{x}$.

All the expressions in the CoM frame are parametrized by the total mass and the symmetric mass ratio, respectively
\begin{equation}
    M=m_1\, +m_2, \qquad \nu=\,\frac{m_1 m_2}{M^2},
\end{equation}
and we also define the quantity $ X_{12}=\sqrt{1-4\nu}$ to simplify some expressions. 

We often make use of the mass-rescaled charges $\eta_A=q_A/m_A$.

We use a mostly plus signature, with the Minkowski metric given by $\eta_{\mu\nu}=(-,+,+,+)$. 

We adopt the multi-index notation $L\equiv i_1\cdots i_\ell$, where $L$ denotes a string of $\ell$ spatial indices. Accordingly, the product of $\ell$ spatial components of a vector $\mathbf{x}$ is written as
$x_L\equiv x_{i_1}\cdots x_{i_\ell}$.


We denote the STF projection in two ways, either using the hat notation $\hat{x}_L \equiv {\rm STF}(x_L) $ or with angular brackets surrounding the indices, e.g.~$x_{\left \langle ij\right \rangle}= x_i x_j\, -\frac{1}{3}x^2 \delta_{ij}$, $x_{ \langle i }v_{ j\rangle}= \frac{1}{2}x_i v_j\, + \frac{1}{2}x_j v_i\, -\frac{1}{3} (\mathbf{x}\cdot \mathbf{v}) \delta_{ij}$, where $ \delta_{ij}$ is the Kronecker delta. As usual $\varepsilon_{ijk}$ stands for the Levi-Civita symbol.

Quantities enclosed in parentheses, such as $(\cdots)_A$, are understood to be evaluated at the position of particle $A$, namely in the limit $\mathbf{x}\to\mathbf{y}_A(t)$. Dirac delta distributions centered on the particle positions are denoted by
$\delta_A(\mathbf{x})\equiv\delta^{(3)}(\mathbf{x}-\mathbf{y}_A)$.
For point-particle sources, ultraviolet divergences arise at the particle positions; these are handled by regularization prescriptions, such as the Hadamard regularization~\cite{HadamardReg,Blanchet:1998vx}.

We also introduce radiative coordinates $(T,\mathbf{X})$, with $R\equiv|\mathbf{X}|$ denoting the distance between the source and the observer, $\mathbf{N}\equiv\mathbf{X}/R$ the unit vector pointing from the source to the observer, and $T_R=T-R/c$ the corresponding retarded time. 

Parentheses on superscripts denote time derivatives with respect to the argument of the quantity; for instance,
$I_{ij}^{(4)}(T_R)\equiv \left.d^4 I_{ij}(t)/dt^4\right|_{t=T_R}$.

Throughout this work we use Gaussian units, in which $G=c=1$, and consequently $\mu_0=\, (\varepsilon_0)^{-1}=4\pi $. Occasionally we retain explicit factors of $c$ and $G$, to cleanly distinguish, respectively, different PN and PM orders.

\section{Setup}
\label{sec: setup}

\subsection{Total 2PN Action}
\label{sec:action2PN}

The total action for the Einstein-Maxwell theory is obtained by 
\begin{equation}
   \label{eq: totalActionSum}
   S=\, S_g\, +S_{\rm EM}\, +S_{m}, 
\end{equation}
where $S_g$ is the Einstein-Hilbert gravitational action, $S_{\rm EM}$ is the EM action and $S_m$ is the matter action.\footnote{We adopt the metric signature $(-,+,+,+)$; consequently, the overall sign of the action differs from that used in our previous work~\cite{Placidi:2025xyi}.} 

The Einstein-Hilbert action in harmonic gauge can be written as
\begin{equation}
\label{eq: Einstein-Hilbert general action}
    S_g\,=\, \frac{c^4}{16 \pi G}\int dt d^3\mathbf{x}\sqrt{-g}g^{\mu \nu}\Bigl(\Gamma^\rho_{\mu \lambda}\Gamma^\lambda_{\nu \rho}\, -\Gamma^\rho_{\mu \nu}\Gamma^{\lambda}_{\rho \lambda}\Bigr),
\end{equation}
where the Christoffel symbols are
\begin{equation}
    \label{eq: Christoffelsymbol}
    \Gamma^{\mu}_{\nu \lambda}=\frac{1}{2}g^{\mu \rho}\Bigl( \partial_\lambda g_{\rho \lambda}+ \partial_\nu g_{\rho \lambda}-\partial_\rho
g_{\nu \lambda}\Bigr).
\end{equation}
Within the PN expansion, the metric can be written as a parametrized expansion in $v/c$~\cite{Damour:1991yw},\footnote{We display only the terms that contribute at relative 2PN order in each component of the metric, as these are sufficient for the computations presented in this paper.}
\begin{subequations}
\label{eq: ExpandedMetric}
    \begin{align}
        g_{00}&=\,-1+ \frac{2}{c^2}V\, -\frac{2}{c^4}V^2\,  + \mathcal{O}(1/{c^6}), \\ 
        g_{0i}&=\, -\frac{4}{c^3}V_i\,+ \mathcal{O}(1/{c^5}), \\ 
        g_{ij}&=\, \delta_{ij}\Biggl(1\,+ \frac{2}{c^2}V\, +\frac{2}{c^4}V^2 \Biggr)\, +\frac{4}{c^4}W_{ij}\, + \mathcal{O}(1/{c^6}),
    \end{align}
\end{subequations}
where $V$, $V_i$, $W_{ij}$ are retarded potentials defined in terms of the stress-energy tensor of matter and such that $V \sim \mathcal{O}(1/c^2)$, $V_i \sim \mathcal{O}(1/c^3)$, and $W_{ij} \sim \mathcal{O}(1/c^4)$. 

To simplify the action~\eqref{eq: Einstein-Hilbert general action} after the replacement of Eq.~\eqref{eq: ExpandedMetric}, we use the Bianchi identity
\begin{equation}
    \label{eq:BianchiIdentity}
    \partial_t V_i\, +\partial_j \Bigl( W_{ij}\, -\frac{1}{2}\delta_{ij} W_{kk}\Bigr)= \mathcal{O}(1/c^2),
\end{equation}
and the harmonic gauge condition
\begin{equation}
\label{eq: HarmonicGauge}
    g^{\mu \nu}\Gamma^\lambda_{\mu \nu}=0. 
\end{equation}


The EM action reads
\begin{equation}
    \label{eq: EM action}
    S_{\rm EM}\, =\, -\frac{1}{16 \pi}\int dt\, d^3\mathbf{x}\, \sqrt{-g}\, F_{\mu \nu}F^{\mu \nu}, 
\end{equation}
where $F_{\mu \nu}$ is the EM field defined in terms of the covariant EM potential $A_{\mu}=(A_0, A_i)$ in the following way $F_{\mu \nu}=\partial_\mu A_{\nu}-\partial_\nu A_\mu$.

Notably, for the PN expansion of Eq.~\eqref{eq: EM action} one should take into account that $A_0 \sim \mathcal{O}(1)$, $A_i \sim \mathcal{O}(1/c)$ and 
\begin{equation}
\label{eq: detg}
\sqrt{-g}=1+ \frac{2V}{c^2}\, +\frac{2}{c^4}(V^2\, + W_{ii}).
\end{equation}
Moreover, simplifications can be made by imposing the Lorenz gauge condition $\partial_\mu A^\mu=0$, which, at the 2PN accuracy we consider, reads 
\begin{equation}
    \label{eq: LorenzGaugePN}
    \partial_i A_i = \partial_t A_0\, +\frac{4}{c^2} (V\partial_t A_0+V_i\partial_iA_0).
\end{equation}
In parallel, another useful relation to simplify the action is 
\begin{equation} \label{eq:partial_der_simplification}
    \partial_i Z\, \partial_i Y = \frac{1}{2} \Bigl[\partial_i^2 ( Z Y)\, - Z\partial_i^2Y\, -Y\partial_i^2Z \Bigr], 
\end{equation}
with $Z$ and $Y$ representing two generic potentials.

Finally, the matter action for minimally charged point particles is given by 
\begin{align}
    \label{eq: MatterAction}
    S_{m}=\, - \sum_{A}\int dt \Biggl[&m_A c^2 \sqrt{g_{\mu \nu} v^\mu_A v^\nu_A/c^2}\, \\ \notag
    &-  q_A A_\mu v^\mu_A  \Biggr].
\end{align}
%

\subsection{The field equations}
\label{subsec:fieldEquation}
Starting from the total action \eqref{eq: totalActionSum}, we can derive the field equations of the problem by varying it with respect to $g_{\mu \nu}$ and $A_\mu$. Introducing the perturbation $h^{\mu\nu}\equiv \sqrt{-g}\,g^{\mu\nu}
-
\eta^{\mu\nu}$, we have
\begin{subequations}
    \begin{equation}
    \label{eq: gravitational_field_equation}
       \Box h^{\mu \nu}=\, \frac{16 \pi G}{c^4} \, \tau^{\mu \nu}, 
    \end{equation}
        \begin{equation}
         \label{eq: EM_field_equation}
       \Box A^{\mu}=\,- \frac{4 \pi}{c} \, \mathcal{J}^{\mu},
    \end{equation}
\end{subequations}
where $\Box= \eta^{\mu \nu}\partial_\mu \partial_\nu=- \partial_0\,+ \nabla^2$ is the d'Alambertian operator in flat space. 

In Eq.~\eqref{eq: gravitational_field_equation} we introduced the stress-energy pseudo-tensor $\tau^{\mu \nu}$, which can be defined as a combination of the matter stress-energy tensor $T^{\mu \nu}$ and the nonlinear interaction term $\Lambda^{\mu \nu}$~\cite{Blanchet:2001aw}
\begin{equation}
\label{eq: pseudo-tensor-stress-energy}
    \tau^{\mu \nu}=\,|g| T^{\mu \nu}+\, \frac{c^4}{16 \pi G} \Lambda^{\mu \nu}.
\end{equation}

For minimally-charged point particles, the expression of $T^{\mu \nu}$ is not modified by the presence of charge by the presence of electric charge, and therefore it is given by the usual expression\footnote{At the level of the action, the fact that in our case there are no additional contributions to $T^{\mu\nu}$ can be understood by noting that the minimal charge coupling in the second line of Eq.~\eqref{eq: MatterAction} has no $g_{\mu\nu}$.}
\begin{equation}
    T^{\mu\nu}
=
\sum_A
\frac{m_A c}{\sqrt{-g}}\,
\frac{v_A^\mu v_A^\nu}{
\sqrt{
-\bigl(g_{\rho\sigma}\bigr)_A \, v_A^\rho v_A^\sigma
}
}
\,
\delta_A(\mathbf{x}).
\end{equation}
The non-linear interaction terms, instead, can be organized according to the split $\Lambda^{\mu \nu} = \Lambda^{\mu \nu}_{\rm GR} +   \Lambda^{\mu \nu}_{\rm EM}$, where the first term collects all the gravitational self-interaction contributions, extensively studied in the case of GWs emitted by neutral binaries~\cite{Blanchet:2001aw}, while the second term includes the interactions between the metric and the EM field. Explicitly, we have \cite{Henry:2023guc,Henry:2023len}
\begin{equation}
    \label{eq: LambdaEM}
    \Lambda^{\mu \nu}_{\rm EM}=4|g|\bigg(F^{\mu \lambda}F^{\nu}_{\, \,\lambda}\,-\frac{1}{4}g^{\mu \nu}\, F^{\alpha\beta}F_{\alpha\beta}\bigg).
\end{equation}
For later convenience, we also introduce the quantities
\begin{equation}
    \label{eq: Sigma_tau}
    \Sigma= \frac{\Bar{\tau}^{00}+\Bar{\tau}^{ii}}{c^2}, \quad \Sigma^i=\frac{\Bar{\tau}^{0i}}{c}, \quad \Sigma^{ij}=\Bar{\tau}^{ij}.
\end{equation}
where the overline denotes the formal PN expansion of the stress-energy pseudotensor. Indeed, the GR+EM decomposition of $\Lambda^{\mu\nu}$ induces an analogous split in the three source densities of Eq.~\eqref{eq: Sigma_tau}. The resulting expressions contain the standard GR contributions, given for instance in Refs.~\cite{Blanchet:1995fg,Blanchet:2001aw}, together with new terms associated with the EM field, whose explicit form is provided in Sec.~\ref{sec:MassCurrentMultipoles}.

Similarly, from the components of $T^{\mu\nu}$ we can define~\cite{Blanchet:1989ki} 
\begin{subequations}
\label{eq: mass_current_stress_densities}
    \begin{align}
        \label{eq: mass_density}
        &\sigma(t, \mathbf{x})\equiv \, \dfrac{T^{00}(t, \mathbf{x})\, +T^{ii}(t, \mathbf{x})}{c^2}, \\
        \label{eq: current_density}
        &\sigma^i(t, \mathbf{x})\equiv \, \dfrac{T^{0i}(t, \mathbf{x})}{c}, \\
        \label{eq: stress_density}
        &\sigma^{ij}(t, \mathbf{x})\equiv \, T^{ij}(t, \mathbf{x}),
    \end{align}
\end{subequations}
respectively the mass density, the current density and the stress density. 
Through these densities, one can write the explicit definitions of the retarded potentials~\cite{Blanchet:1998vx} that parametrize the metric in Eq.~\eqref{eq: ExpandedMetric}, e.g.
\begin{equation}
    V(\mathbf{x},t)
\equiv
G \int \frac{d^3\mathbf{y}}{|\mathbf{x}-\mathbf{y}|}
\,\sigma\!\Bigl(\mathbf{y},t-\frac{|\mathbf{x}-\mathbf{y}|}{c}\Bigr).
\end{equation}

For our computation, it is useful to expand the retardation effects entering these potentials in PN orders, which results in the definition of associated instantaneous potentials~\cite{Blanchet:1995fg}.

In particular, at 2PN accuracy, we have~\cite{Blanchet:2001aw,Blanchet:1998vx} 
\begin{align}
\label{eq: retarded_potentials}
    &V\equiv U+\frac{1}{2c^2}\partial_t^2 X+\mathcal{O}(1/{c^4}), \quad V_i = U_i+\mathcal{O}(1/{c^2}), \cr
   &W_{ij} = U_{ij} + \mathcal{O}(1/c),
\end{align}
where 
\begin{subequations}
\label{eq: Newtonina_potentials}
    \begin{align}
        \label{eq: U}
        &U=\, \int \frac{d\mathbf{z}}{|\mathbf{x}-\mathbf{z}|}\, \sigma(t,\mathbf{z}), \\
        \label{eq: Ui}
        &U_i=\, \int \frac{d\mathbf{z}}{|\mathbf{x}-\mathbf{z}|}\, \sigma_i(t,\mathbf{z}), \\
        \notag
        &U_{ij}=-\int \frac{d\mathbf{z}}{|\mathbf{x}-\mathbf{z}|}\, \Biggl[\sigma_{ij}-\delta_{ij}\sigma^{ss}\\ \label{eq: Uij}
        &\quad+\frac{1}{4\pi} \left(\partial_i U \partial_jU-\partial_i A_0\partial_jA_0\right)\Biggr](t,\mathbf{z}), \\
        \label{eq: X}
        &X=\, \int d\mathbf{z}|\mathbf{x}-\mathbf{z}|\, \sigma(t,\mathbf{z}).
    \end{align}
\end{subequations}
Regarding the EM field equation \eqref{eq: EM_field_equation}, the source is given by the vector current
\begin{equation}
    \mathcal{J}^{\mu} = J^\mu
-
\frac{c}{4\pi}\,\Phi^\mu,
\end{equation}
where we isolated the compact-support term
\begin{equation} \label{eq:compact_J}
   J^\mu= \sum_A
\frac{q_A v_A^\mu}{\sqrt{-g}}\,
\delta_A(\mathbf{x}),
\end{equation}
and the non-linear interaction term
\begin{align}
\label{eq:interaction_J}
&\Phi^\mu
=
-h^{\alpha\beta}\partial_\alpha\partial_\beta A^\mu
-\partial^\mu h^{\alpha\beta}\partial_\alpha A_\beta
-F_{\alpha\beta}\partial^\alpha h^{\beta\mu}
\nonumber\\
&\quad-\frac{1}{2}F^\mu{}_{\alpha}\partial^\alpha h
+h^\mu{}_{\nu}F_{\alpha\beta}
  \partial^\alpha h^{\beta\nu}
-h^{\alpha\lambda}F_{\alpha\beta}
  \partial_\lambda h^{\beta\mu}
\nonumber\\
&\quad-\frac{1}{2}h^{\alpha\beta}F^\mu{}_{\alpha}
  \partial_\beta h
+\frac{1}{2}F^\mu{}_{\alpha}h_{\lambda\rho}
  \partial^\alpha h^{\lambda\rho}
+O(h^3A),
\end{align}

For our purposes, it is convenient to split the total vector current as $\mathcal{J}^{\mu}=( c\, \rho, \mathcal{J}^i)$, and thus write two distinct field equations for $A_0$ and $A_i$, i.e.
\begin{subequations}
\label{eq: A0-Ai-field_equations}
    \begin{align}
        \label{eq: field_equation_A0}
        \Box A_0\,=&\,4\pi \rho, \\
        \label{eq: field_equation_Ai}
        \Box A_i\,=&-\frac{4\pi}{c} \mathcal{J}_i.
    \end{align}
\end{subequations}
The explicit PN-expanded form of these sources can be obtained from the corresponding PN expansion of Eqs.~\eqref{eq:compact_J} and \eqref{eq:interaction_J}.
To compute the 2PN energy flux at infinity, we only need the leading $1/R$ components of the sources, with $\rho$ retained at relative 2PN accuracy and $\mathcal{J}_i$ at relative 1PN accuracy. Using Eqs.~\eqref{eq:BianchiIdentity},~\eqref{eq: LorenzGaugePN}, and \eqref{eq:partial_der_simplification}, 
we obtain 
\begin{widetext}
    \begin{subequations}
    \label{eq: EM_source_functions}
    \begin{align}
        \notag
        &\rho=q_1 \delta_1(\mathbf{x})\, +q_2 \delta_2(\mathbf{x})\,-\frac{1}{2 \pi c^2}\partial_i (V \partial_i A_0)+\frac{1}{4\pi c^4}\Biggl\{4 \partial_t (V \partial_t A_0)\\ \label{eq: rho}
        &+\, \partial_i \Bigl[4 W_{ij}\partial_j A_0+2V\partial_tA_i-2(V^2+W_{kk})\partial_i A_0 +4V_j \partial_jA_i\, -4V_j \partial_i A_j\Bigr]\Biggr\}\, + \mathcal{O}(1/c^6), \\ \notag
        &\mathcal{J}_i= \Bigl(q_1 \delta_1(\mathbf{x})\,+q_2 \delta_2(\mathbf{x})\Bigr)v_i\, +\frac{1}{4\pi c^2}\Biggl[ U \partial_j^2 A_i\, -A_i \partial_j^2U\,+2 U_i \partial_j^2 A_0\, -2A_0\partial_j^2U_i\, \\ 
        \label{eq:Ji_EM} 
        &\quad-4U\partial_i \partial_jA_j\, -4U_j \partial_i \partial_j A_0\, +2A_j \partial_i \partial_j U\, +2 A_0 \partial_i \partial_j U_j\Biggr]\, + \mathcal{O}(1/c^4).
    \end{align}
\end{subequations}
\end{widetext}
At 1PN order, Ref.~\cite{Khalil:2018aaj} found that the electric charge density reduces to its flat-space counterpart,
$\rho_e\equiv\sum_A q_A\delta_A(\mathbf{x})$. We instead find a nonvanishing 1PN correction to $\rho_e$. The origin of this difference
appears to be a cancellation occurring in the derivation of Ref.~\cite{Khalil:2018aaj} which we do not encounter in the present calculation.
As discussed below, the additional term found here is independently supported by the agreement with the EFT computation and, at the level of the 1PN energy flux, by the agreement with the results of Ref.~\cite{Julie:2018lfp}.
The 1PN term in the current density $\mathcal{J}_i$ and the 2PN term in the charge density $\rho$ are new to this work.

To obtain the explicit expression for the NLO of the EM source in Eqs.~\eqref{eq: EM_source_functions}, we consider the NLO solution of the $A_0$ field equation, and the LO solution of the $A_i$ field equations~\eqref{eq: EM_field_equation}, namely
\begin{widetext}
\begin{subequations}
 \label{eq: LO_A_0_A_i}
   \begin{align}
  \notag
    A_0=&\, -\frac{\eta_1\, m_1}{|\mathbf{x}-\mathbf{y}_1|}-\frac{\eta_2\, m_2}{|\mathbf{x}-\mathbf{y}_2|}\,\\ \notag
    &+\frac{1}{c^2}\Biggl\{\left(\frac{\eta_1 m_1^2}{|\mathbf{x}-\mathbf{x}_1|}+\frac{\eta_2 m_2^2}{|\mathbf{x}-\mathbf{x}_2|} \right)+\left(\frac{m_1}{|\mathbf{x}-\mathbf{x}_1|}+\frac{m_2}{|\mathbf{x}-\mathbf{x}_2|} \right) \left(\frac{\eta_1 m_1}{|\mathbf{x}-\mathbf{x}_1|}+\frac{\eta_2 m_2}{|\mathbf{x}-\mathbf{x}_2|} \right)\\ \notag
    &-\frac{\eta_2 m_2 m_1}{|\mathbf{x}-\mathbf{x}_1|}-\frac{\eta_1 m_1 m_2}{|\mathbf{x}-\mathbf{x}_2|} \, -\frac{\eta_1 m_1}{2}\left( \frac{v_1^2}{|\mathbf{x}-\mathbf{x}_1|}-\mathbf{n}_1\cdot \mathbf{a}_1-\frac{(\mathbf{n}_1\cdot \mathbf{v}_1)^2}{|\mathbf{x}-\mathbf{x}_1|}\right)\\  \label{eq: NLO_A0}
    &-\frac{\eta_2 m_2}{2}\left( \frac{v_2^2}{|\mathbf{x}-\mathbf{x}_1|}-\mathbf{n}_2\cdot \mathbf{a}_2-\frac{(\mathbf{n}_2\cdot \mathbf{v}_2)^2}{|\mathbf{x}-\mathbf{x}_2|}\right)
     \Biggr\}+\mathcal{O}\left(1/c^3\right) \\ \label{eq: leading Ai}
    A_i=&\, \frac{1}{c}\Biggl[\frac{\eta_1\, m_1}{|\mathbf{x}-\mathbf{y}_1|}(v_i)_1\,+\frac{\eta_2\, m_2}{|\mathbf{x}-\mathbf{y}_2|}(v_i)_2\Biggr]\, +\mathcal{O}\left(1/c^2\right).
\end{align} 
\end{subequations}
together with the expression for the metric potentials discussed above.
%
\subsection{Circular limit}
\label{subsec: CoM_circular_limit}

As anticipated, we will specialize our explicit results to binaries evolving along circular orbits. In the CoM frame,\footnote{Throughout this paper, we use the 2PN transformation to the CoM frame provided in Ref.~\cite{Placidi:2025xyi}.} the conservative sector of the 2PN equations of motion admits circular periodic solutions characterized by the orbital frequency $\omega$, for which
\begin{equation}
    \label{eq: EoM_circular_orbits}
    \mathbf{a}_{\rm cons}=-\omega^2 \mathbf{r}\,.
\end{equation}
From the 2PN results of Ref.~\cite{Placidi:2025xyi}, we have
\begin{align}
 \notag
\omega^2
&=\frac{M(1-\eta_1\eta_2)}{r^3}
-\frac{M^2}{c^2r^4}\Biggl\{
3-\nu
-\frac{9}{2}\eta_1\eta_2
+2\eta_1\eta_2\nu
+\frac{\eta_2^2}{2}
+\frac{\eta_1^2}{2}
+\eta_1^2\eta_2^2\Bigl(\frac12-\nu\Bigr)
+X_{12}\Bigl(
\frac{\eta_1^2}{2}
-\frac{\eta_2^2}{2}
\Bigr)
\Biggr\}
\\ \notag
&+\frac{M^3}{c^4r^5}\Biggl\{
6+\frac{41}{4}\nu+\nu^2
-\frac{87}{8}\eta_1\eta_2
-\frac{99}{4}\eta_1\eta_2\nu
-3\eta_1\eta_2\nu^2
+2\eta_2^2+\frac14\eta_2^2\nu
+2\eta_1^2+\frac14\eta_1^2\nu
\\ \notag
&\qquad
-\frac12\eta_1\eta_2^3
+\frac54\eta_1\eta_2^3\nu
-\frac12\eta_1^3\eta_2
+\frac54\eta_1^3\eta_2\nu
+\eta_1^2\eta_2^2
\Bigl(
2+\frac{49}{4}\nu+3\nu^2
\Bigr)
-\eta_1^3\eta_2^3
\Bigl(
\frac18+\frac34\nu+\nu^2
\Bigr)
\\
\label{eq:orbital_frequency_circular_orbit}
&\qquad
+X_{12}\Biggl(
-2\eta_2^2
+\frac54\eta_2^2\nu
+2\eta_1^2
-\frac54\eta_1^2\nu
+\frac12\eta_1\eta_2^3
-\frac54\eta_1\eta_2^3\nu
-\frac12\eta_1^3\eta_2
+\frac54\eta_1^3\eta_2\nu
\Biggr)
\Biggr\}
+\mathcal{O}(1/c^6).
\end{align}
\end{widetext}

The orbital frequency naturally defines the gauge-invariant PN parameter
\begin{equation}
\label{eq: x_definition}
x=\left(\frac{M\omega}{c^3}\right)^{2/3}.
\end{equation}
This parameter will be used to express the energy flux and the waveform modes in the circular limit, which is taken by imposing $\mathbf{n}\cdot\mathbf{v}=0$ and $|\mathbf{v}|=\omega \, r$, in compliance with Eq.~\eqref{eq: EoM_circular_orbits}.

\subsection{Complementary EFT computation}
\label{subsec:PN_EFT_radiation}
The same Einstein-Maxwell action provides the starting point for the independent radiation calculation in effective field theory, extending the formalism presented in \cite{Goldberger:2004jt, Goldberger:2009qd, Ross:2012fc,Leibovich:2019cxo,Amalberti:2023ohj, Amalberti:2024jaa, Mandal:2024iug}.
Using the method of regions, we separate both the metric and the four-vector into potential and radiation modes, where the former scale as $(k^0,k^i) \sim \left(\frac{v}{c r},\frac{1}{r}\right)$, whereas the latter scale as $(k^0,k^i) \sim \left(\frac{v}{c r},\frac{v}{c r}\right)$.
To perform the computations, we use background covariant gauge fixing, imposed using the background field method~\cite{Goldberger:2004jt,Porto:2016pyg,Amalberti:2024jaa,Mandal:2024iug}. This preserves the gauge invariance of the radiation effective theory after the potential modes are integrated out, and in particular ensures that the Ward identities shall hold, providing a non-trivial check for our computations. 
Specifically, we employ background-covariant de Donder gauge for the potential graviton and background-covariant Lorenz gauge for the potential photon.

We compute the one-point radiation function by integrating out the potential modes in diagrams with one external radiation field. We then extract the electromagnetic and gravitational source multipoles of the binary system from the one-point function using the construction of Ref.~\cite{Ross:2012fc,Amalberti:2023ohj}, further evaluating the instantaneous energy flux to 2PN order relative to the leading electromagnetic dipole contribution.

Concerning the transformation connecting the gauge used in this EFT formulation and the harmonic gauge used in the rest of the paper, the corresponding sets of equations of motion are related by a contact transformation of the particle positions, starting at the 2PN order. Explicitly we have
\begin{align} y_{1,\mathrm{H}}^i &= y_{1,\mathrm{EFT}}^i +\frac{1}{c^4}\,\xi_{1,\mathrm{EFT}}^i +\mathcal{O}(1/c^6), \\ y_{2,\mathrm{H}}^i &= y_{2,\mathrm{EFT}}^i +\frac{1}{c^4}\,\xi_{2,\mathrm{EFT}}^i +\mathcal{O}(1/c^6), \end{align}
where
\begin{align} \xi_{1,\mathrm{EFT}}^i &= 2G^2 M m_2\, \frac{n_{\mathrm{EFT}}^i}{r_{\mathrm{EFT}}}, \\ \xi_{2,\mathrm{EFT}}^i &= -2G^2 M m_1\, \frac{n_{\mathrm{EFT}}^i}{r_{\mathrm{EFT}}}.
\end{align}
The relative separation transforms therefore as
\begin{equation} \label{eq:guage_EFT_to_H}
r_{\mathrm{H}}^i = r_{\mathrm{EFT}}^i \bigg( 1+\frac{2G^2M^2}{c^4 r_{\mathrm{EFT}}^2} \bigg) +\mathcal{O}(1/c^6). \end{equation}

For notational simplicity, in the remainder of the text we drop the label
``H'' from all dynamical variables, with the understanding that all quantities
are henceforth expressed in the harmonic coordinate system.

\section{Radiative multipole moments of the vector field}
\label{sec:ElectricMagneticMultipoles}

In this section, we derive the expressions for the STF multipole moments that describe the vector field $A_\mu$ at infinity.
Following Refs.~\cite{Khalil:2018aaj,Henry:2023len} and employing radiative coordinates, we specifically refer to the multipolar decomposition of the EM field in the transverse basis defined by $A^T_0=N_iA^T_i=0$, that is

\begin{align}
\label{eq:AT}
    &A_i^{T}(T,\mathbf X)=\frac{1}{R}
P_{ij}(\mathbf N)
\sum_{\ell\geq1}
\frac{1}{c^\ell\ell!}
\bigg[
N_{L-1}\mathcal E_{jL-1}(T_R)
\cr&\quad-
\frac{\ell}{c(\ell+1)}
\epsilon_{jab}N_{aL-1}\mathcal B_{bL-1}(T_R)
\bigg]
+\mathcal{O}(1/R^{2}) ,
\end{align}
where $P_{ij}(\mathbf N)\equiv \delta_{ij}-N_iN_j$ is the transverse projector.

The goal of this section is therefore to derive explicit expressions, with overall 2PN accuracy, for the electric moments $\mathcal{E}_L$ and the magnetic moments $\mathcal{B}_L$. 

Within the MPM-PN matched formalism, the radiative moments can be decomposed as~\cite{Henry:2023len}
\begin{align}
\label{eq: definition_electric_radiative_momemnts}
    \mathcal{E}_L=& Q_L^{(\ell)}+\delta \mathcal{E}_L^{\rm tail}\, +\mathcal{O}(1/c^5), \\ \label{eq: definition_magnetic_radiative_momemnts}
        \mathcal{B}_L=& M_L^{(\ell)}+\delta \mathcal{B}_L^{\rm tail}\, +\mathcal{O}(1/c^5),
\end{align}
where $Q_L$ and $M_L$ are the source electric and magnetic multipole moments, respectively. These encode the linear contribution to the radiation field, while the additional terms account for nonlinear propagation effects in the curved spacetime generated by the source. In particular, the tail contributions $\delta \mathcal{E}_L^{\rm tail}$ and $\delta \mathcal{B}_L^{\rm tail}$ arise from the scattering of the emitted radiation off the background monopolar field and are the only nonlinear terms entering at the relative 2PN order.

In the following subsection, we discuss separately the two types of contributions described above.

\subsection{Source multipole moments of the vector field}

The source moments $Q_L$ and $M_L$ are fixed by the standard matching procedure between the PN-expanded near-zone solution and the MPM expansion in the exterior region. This construction relies on the existence of an overlap region where both expansions are valid, and allows one to express the irreducible STF moments of the exterior vector field in terms of the effective source functions entering the vector wave equation \cite{Blanchet:2013haa}. In the present charged-binary problem, the same assumptions underlying the usual MPM-PN matching can be applied to the EM sector, as discussed in Ref.~\cite{Henry:2023len}. 

The moments $Q_L$ and $M_L$ can therefore be directly connected to the effective source functions $\rho$ and $\mathcal{J}_i$ defined in Eqs.~\eqref{eq: EM_source_functions}. At the accuracy required for the 2PN computation of the EM energy flux at infinity, we have
\begin{align}
    \notag
        &Q_L={\rm FP}_{B=0}  \int d^3\mathbf{x} \Biggl[\hat{x}_L \rho\, +\frac{x^2 \hat{x}_L}{2(2\ell+3)c^2}\frac{d^2 \rho}{dt^2}\,\\ \notag
        &\quad+\frac{x^4 \hat{x}_L}{8(2\ell+3)(2\ell+5)c^4}\frac{d^4\rho}{dt^4}\, -\frac{(2\ell+1)\hat{x}_{aL}}{(\ell+1)(2\ell+3)c^2}\frac{d\mathcal{J}_a}{dt}\,\\
        \label{eq: QL_general}
        &\quad-\frac{(2\ell+1)\, x^2\,\hat{x}_{aL}}{2(\ell+1)(2\ell+3)(2\ell+5)c^4}\frac{d^3\mathcal{J}_a}{dt^3} \Biggr]\, +\mathcal{O}(1/c^6),\\ 
       \notag
        &M_L={\rm FP}_{B=0} \varepsilon_{ab\langle i_\ell}\int d^3\mathbf{x} \Biggl[x_{L-1\rangle b}\, \mathcal{J}_a\, \\ 
         \label{eq: ML_general}
         &\quad+\frac{x^2x_{L-1\rangle b}}{2(2\ell+3)c^2}\frac{d^2 \mathcal{J}_a}{dt^2} \Biggr]\, +\mathcal{O}(1/c^4).
    \end{align}
Because of the non-compact support of the source, the integrals become divergent at spatial infinity, i.e. in the limit $|\mathbf{x}|\rightarrow + \infty$. To regularize these divergences, we employ the finite part operation based on analytic continuation, denoted by $\rm FP_{B=0}$, where $B$ is a complex number (see Refs.~\cite{Blanchet:1998in, Blanchet:2001aw} for details). This prescription is equivalent to the Hadamard \emph{partie finie} regularization~\cite{HadamardReg,Blanchet:2000nu}. 

For the next-to-next-to-leading-order (N${}^2$LO) energy flux, the electric dipole $Q_i$ is needed at relative 2PN accuracy, the magnetic dipole $M_i$ and electric quadrupole $Q_{ij}$ at relative 1PN accuracy, and the magnetic quadrupole $M_{ij}$ and electric octupole $Q_{ijk}$ at Newtonian order.

Following Ref.~\cite{Blanchet:1995fg}, we decompose the contributions to $Q_L$ and $M_L$ into three classes, which we refer to as \textit{compact}, \textit{quadratic}, and \textit{cubic} terms. The \textit{compact} terms are those for which no finite-part prescription is required, since the spatial integrals are restricted to the compact support of the material source. They contain explicit source contributions, namely Dirac delta distributions $\delta_A(\mathbf{x})$ and their derivatives, and coincide with the terms obtained by solving the linearized problem at 2PN order. The \textit{quadratic} terms collect all contributions involving integrals of products of two Newtonian-like potentials, or of their spatial derivatives. For these terms, time derivatives can be factorized outside the integrals and act on the finite part of the resulting three-dimensional spatial integrals. Finally, the \textit{cubic} terms involve integrals of quantities that are trilinear in the source variables.

While the compact terms can be evaluated straightforwardly from the Dirac delta distributions and their derivatives, the quadratic and cubic terms require the finite-part techniques developed to handle analogous non-compact integrals in the standard MPM-PN framework. Further details can be found, for instance, in Refs.~\cite{Blanchet:1995fg,Blanchet:2001aw}.

%
The resulting circular-orbit expressions for the electric and magnetic source moments required at 2PN order read

\begin{widetext}
    \begin{subequations}
    \label{eq: electric_magnetic_multipoles_moments_explicit}
        \begin{align}
          \notag Q_i=&M\, \nu\, \left(\eta_1- \eta_2\right)\, r_i\, +\frac{M^2\, \nu\, r_i }{10 r\, c^2}\Biggl\{X_{12}\left(1- \eta_1 \eta_2\right)\left(\eta_1+\eta_2\right)+\eta_1 \eta_2 (1-2 \nu )-11+2 \nu\Biggr\} \\ \notag
          &+\frac{M^3\,\nu\, r_i}{240\, r^2\, c^4}\Biggl\{X_{12}(\eta_1+\eta_2)\Bigl[319+62 \nu +28 \eta_2^2 (5+3 \nu )-4 \eta_1 \eta_2 (177+73 \nu )+\eta_1^2 \bigl(28 (5+3 \nu )+\eta_2^2
   (109+62 \nu )\bigr) \Bigr]   \\ \notag
   &\qquad+(\eta_1-\eta_2)\Bigr[ \eta_1^2 \left(140-84 \nu +\eta_2^2 (171-684 \nu +118
   \nu ^2)\right)-207-4100 \nu +118 \nu ^2+28 \eta_2^2 (5-3
   \nu )\\ \label{eq: Q_i_2PN}
   &\qquad-4 \eta_1 \eta_2 \left(19-1490 \nu +59 \nu
   ^2\right)\Bigr]
   \Biggr\},
   \\ \label{eq: Mi}
          M_i=&\frac{M\, \nu}{2}\varepsilon_{ijk}v^jr^k\Biggl\{X_{12}\left(\eta_1- \eta_2\right)\, -\eta_1\,-\eta_2\,+\frac{M}{3\,r\,c^2}\Bigl[13\,X_{12}\left(\eta_1-\eta_2\right)-\left(\eta_1+\eta_2\right)(13-2\nu) \Bigr]\Biggr\},  \\ \notag
          Q_{ij}=&\frac{M\, \nu\, }{2}\Biggl\{ \hat{r}_{ij}\Bigl[\eta_1\,+\eta_2\, -X_{12}\left(\eta_1-\eta_2\right)\Bigr]\, +
          \frac{5 M r^2\nu}{42\,  c^2}\hat{v}_{ij}\left((1-3\nu)(\eta_1+\eta_2)-X_{12}(\eta_1-\eta_2)(1-\nu)\right)\, \\ \label{eq: Qij}
          &+\frac{M^2\nu}{42\, r\, c^2} \hat{r}_{ij}\left( X_{12}(\eta_1-\eta_2)(47+37\nu+\eta_1 \eta_2(5+37\nu)-5(1-\eta_1\eta_2)(1-3\nu)(\eta_1+\eta_2)\right) \Biggr\},\\ 
          \label{eq: Mij}
          M_{ij}=&\frac{M\, \nu}{2} \varepsilon_{sp\langle i}r_{j \rangle}r_s v_p\Bigl\{X_{12}\left(\eta_1+\eta_2\right)\,-\left(\eta_1-\eta_2\right)(1-2\nu) \Bigr\}, \\ 
          \label{eq: Qijk}
          Q_{ijk}=&\frac{M\, \nu\, \hat{r}_{ijk}}{2}\Bigl\{\left(\eta_1-\eta_2\right)(1-2\nu)\, -X_{12}\left(\eta_1+\eta_2\right) \Bigr\}.
        \end{align}
    \end{subequations}
\end{widetext}
%
The results above coincide with the
corresponding EFT expressions once the latter are expressed in the harmonic
coordinates used here. For \(M_i\), \(Q_{ij}\), \(M_{ij}\), and \(Q_{ijk}\),
the expressions agree directly at the required PN order. The
electric dipole \(Q_i\), being at relative 2PN order, additionally requires the contact transformation of Eq.~\eqref{eq:guage_EFT_to_H}, which cleanly reabsorbs the difference
\begin{equation}
\delta Q_i
\equiv
Q_i^{\mathrm H}-Q_i^{\mathrm{EFT}}
=
-\frac{2G^2M^3\nu}{c^4 r^2}
(\eta_1-\eta_2)\,r_i.
\label{eq:delta_Q_EFT_H}
\end{equation}

\subsection{Nonlinear propagation effects}
\label{sec:nonlinear_propagation_effects}

We now turn to the nonlinear propagation effects entering the vector radiative moments. At the accuracy required for the 2PN EM energy flux, the relevant contribution is just the leading tail correction to the radiative electric dipole. This effect is generated by the quadratic interaction between the mass monopole of the source and the electric-dipole sector of the vector field, schematically $M\times Q_i$, and describes the backscattering of the emitted EM radiation off the long-range monopolar gravitational field.

The formal derivation of this contribution within the MPM-PN approach was carried out in Ref.~\cite{Henry:2023len}. The computation consists in solving the PM-expanded vacuum equation for the vector field, selecting the quadratic interaction between the mass-monopole part of the metric and the linear electric-dipole vector field. After applying the regularized retarded Green function and extracting the leading $1/R$ behavior in radiative coordinates, one obtains~\cite{Henry:2023len}
\begin{equation}
\label{eq:electric_dipole_tail}
\mathcal{E}_i
=
\frac{d Q_i}{dT_R}
+
\delta \mathcal{E}_i^{\rm tail}
+
\mathcal{O}\left(\frac{1}{c^5}\right),
\end{equation}
with
\begin{align}
\label{eq:electric_dipole_tail_integral}
&\delta \mathcal{E}_i^{\rm tail}
=
\frac{2M}{c^3}
\int_0^{+\infty} d\tau
\bigg[
\ln\left(\frac{\tau}{2b_0}\right)+\frac{3}{4}
\bigg]
Q_i^{(3)}(T_R-\tau).
\end{align}
Here $b_0$ is the arbitrary time scale entering the definition of the radiative coordinates. The analogous magnetic-dipole tail has the same structure, with $Q_i$ replaced by $M_i$, but it does not enter the EM flux at the 2PN accuracy considered in the present analysis.

For quasi-circular orbits, the hereditary integral in Eq.~\eqref{eq:electric_dipole_tail_integral} can be evaluated explicitly by decomposing the source moment into Fourier harmonics. Namely, each component of $Q_i$ can be written as a finite sum of terms proportional to $e^{-i n\omega U}$, with integer $n$. Since the tail term is already of relative 1.5PN order, the orbital frequency can be treated as constant inside the hereditary integral. The required integrals are then evaluated mode by mode using
\begin{subequations} \label{eq:tail_int_formulas}
\begin{align}
&\int_0^{+\infty} d\tau\, e^{-i\Omega\tau}
=
-\frac{i}{\Omega},
\\
&\int_0^{+\infty} d\tau\, e^{-i\Omega\tau}
\ln\left(\frac{\tau}{2\tau_0}\right)
=
-\frac{1}{\Omega}
\bigg[
\frac{\pi}{2}\,\mathrm{sgn}(\Omega)
\cr&\quad-i\left(\ln(2|\Omega|\tau_0)+\gamma_{\rm E}\right)
\bigg],
\end{align}
\end{subequations}
for nonzero frequency $\Omega$. This procedure turns the hereditary tail contribution into an explicit circular-orbit expression containing the characteristic $\pi$ terms and logarithms associated with wave propagation in the monopolar background.

\section{Radiative multipole moments of the gravitational field}
\label{sec:MassCurrentMultipoles}
In this section, we compute the charge-dependent corrections to the mass- and current-type moments that determine STF multipolar decomposition of the gravitational radiation field at infinity. 

In radiative coordinates $(T,\mathbf{X})$, the transverse-traceless metric perturbation takes the general form~\cite{Thorne:1980ru}
\begin{align}
\label{eq:TT_projection}
&h_{mn}^{\rm TT}(\mathbf{X},T)
=
\frac{4G}{c^2R}
\mathcal{P}^{\rm TT}_{mnij}(\mathbf{N})
\sum_{\ell=2}^{\infty}
\frac{1}{c^\ell \ell!}
\Biggl[
N_{L-2}U_{ijL-2}(T_R)\cr&\quad
-\frac{2\ell}{c(\ell+1)}
N_{aL-2}\varepsilon_{ab(i}V_{j)bL-2}(T_R)
\Biggr]
+\mathcal{O}(1/R^2).
\end{align}
This equation defines the radiative mass and current moments $U_L$ and $V_L$, respectively, that encode the gravitational radiation measured at infinity. The transverse-traceless projector $\mathcal{P}^{\rm TT}_{mnij}(\mathbf{N})$ is built from the transverse projector introduced below Eq.~\eqref{eq:AT}, and reads
\begin{align}
    \mathcal{P}_{mnij}(\mathbf{N})
    =
    \mathcal{P}_{mi}(\mathbf{N})\mathcal{P}_{nj}(\mathbf{N})
    -\frac{1}{2}\mathcal{P}_{mn}(\mathbf{N})\mathcal{P}_{ij}(\mathbf{N}) .
\end{align}


Within the MPM-PN formalism, The radiative moments $U_L$ and $V_L$ defined by Eq.~\eqref{eq:TT_projection} are obtained from the corresponding STF source moments $I_L$ and $J_L$, which describe the multipolar structure of the source in the near zone. The relation between the two sets of moments at relative 2PN order has the schematic form
\begin{subequations}
\begin{align}
U_L
&=
I_L^{(\ell)}
+\delta U_L^{\rm tail}
+\delta U_L^{\rm mem\, EM}
+\mathcal{O}(1/c^5),
\\
V_L
&=
J_L^{(\ell)}
+\delta V_L^{\rm tail}
+\mathcal{O}(1/c^5).
\end{align}
\end{subequations}
As usual, tail and memory contributions encode nonlinear propagation effects between the source and the observer.

A 2PN-accurate waveform and energy flux require the mass quadrupole $U_{ij}$ through relative 2PN order, the mass octupole $U_{ijk}$ and current quadrupole $V_{ij}$ through relative 1.5PN order, the mass hexadecapole $U_{ijkl}$ and current octupole $V_{ijk}$ through relative 1PN order, and the moments $U_{ijklm}$ and $V_{ijkl}$ at Newtonian order.

The charge-dependent contributions to the radiative multipoles required at this accuracy arise through several distinct mechanisms. Schematically, they arise from:
\begin{itemize}
    \item the order-reduction procedure and the transformation to the center-of-mass frame, which affect both the source moments and the nonlinear contributions;
    \item the EM components of the effective source densities defined in Eq.~\eqref{eq: Sigma_tau};
    \item the new nonlinear memory terms induced by the EM field, that is $\delta U_L^{\rm mem\, EM}$.
\end{itemize}
The remainder of this section is devoted to the explicit evaluation of these contributions.
\subsection{Source multipole moments}
\label{sec:SourceMassCurrentMultipoles}
In the nonlinear theory, closed integral expressions for the STF mass-type and current-type source multipole moments, valid to all orders in the PN expansion, are given in Ref.~\cite{Blanchet:1998in}. In terms of the effective source densities defined in Eq.~\eqref{eq: Sigma_tau}, they read
\begin{widetext}
\begin{subequations}
\label{eq: IL_JL_general_definition}
\begin{align}
    \notag
    I_L(t)=&\, {\rm FP}_{B=0} \int d^3\mathbf{x}|\tilde{\mathbf{x}}|^B\int_{-1}^{1}dz \Biggl\{ \delta_\ell(z) \hat{x}_L\,  \Sigma\,- \frac{4\,(2 \ell+1)}{c^2\, (\ell+1)(2\ell+3)}\delta_{\ell+1}(z) \hat{x}_{iL}\, \partial_t\Sigma_i\, \\ 
    \label{eq: IL_general_definition}
    &+\frac{2(2\ell+1)}{c^4(\ell+1)(\ell+2)(2\ell+5)}\delta_{\ell+2}(z) \hat{x}_{ijL}\, \partial_t^2\Sigma_{ij}\Biggr\}\Bigl(\mathbf{x},t+z\frac{|\mathbf{x}|}{c}\Bigr), \\ \notag
    J_L(t)=&\, {\rm FP}_{B=0} \varepsilon_{ab\langle i_\ell} \int d^3\mathbf{x}|\tilde{\mathbf{x}}|^B\int_{-1}^{1}dz \Biggl\{\delta_\ell(z)  \hat{x}_{L-1\rangle a}\, \Sigma_b\, \\
    \label{eq: JL_general_definition}&-\frac{2\ell+1}{c^2\, (\ell+2)(2\ell+3)}\delta_{\ell+1}(z) \hat{x}_{L-1\rangle ac}\,\partial_t\Sigma_{bc}\Biggr\} \Bigl(\mathbf{x},t+z\frac{|\mathbf{x}|}{c}\Bigr).
\end{align}
\end{subequations}
\end{widetext}
Here $|\tilde{\mathbf{x}}|^B\equiv(|\mathbf{x}|/r_0)^B$ is the analytic regularization factor entering the finite-part prescription ${\rm FP}_{B=0}$, with $r_0$ an arbitrary length scale. This factor is needed to handle the non-compact support of the source densities; the scale $r_0$ drops out of physical observables after all contributions are combined.

We recall that, for point-particle sources, the integrals in Eqs.~\eqref{eq: IL_JL_general_definition} also develop ultraviolet divergences at the particle positions. These divergences are handled by the Hadamard regularization prescription~\cite{HadamardReg,Blanchet:1998vx}, as already mentioned in Sec.~\ref{subsec: notation}.

We decompose the effective source densities $\Sigma$, $\Sigma_i$, and $\Sigma_{ij}$ into their GR and EM contributions, following the split described in Sec.~\ref{subsec:fieldEquation}. The contributions to the source moments arising from the GR part of these densities can be evaluated by applying the same procedure as in Ref.~\cite{Blanchet:1995fg}. In the present case, however, the resulting expressions acquire additional charge-dependent corrections, since the order-reduction procedure and the transformation to the CoM frame must be performed using the charge-dependent 2PN equations of motion and coordinate transformations derived in Ref.~\cite{Placidi:2025xyi}.

Further contributions come from the explicit EM parts of the effective source densities. These are generated both by the EM stress-energy tensor and by the EM nonlinearities contained in $\Lambda^{\mu\nu}$. Substituting these terms into the general definitions~\eqref{eq: IL_JL_general_definition}, and expanding the retardations to the PN accuracy required here, gives the additional EM pieces
\begin{widetext}
\begin{align}
 \notag
    I_L^{\rm EM}=\, &{\rm FP}_{B=0} \int d^3 \mathbf{x}|\mathbf{x}|^B \Biggl\{ \hat{x}_L \Biggl[ \Sigma^{\rm EM}\,+\frac{|\mathbf{x}|^2}{2(2\ell+3) c^2}\partial_t^2 \Sigma^{\rm EM}\, +\frac{|\mathbf{x}|^4}{8(2\ell+3)(2\ell+5) c^4}\partial_t^4 \Sigma^{\rm EM}\Biggr] \\ 
       \label{eq: EM_mass_multipole_moments}
       &\qquad -\frac{4(2\ell+1)\, \hat{x}_{iL}}{(\ell+1)(2\ell+3) c^2}\partial_t\Sigma_i^{\rm EM}\, + \frac{2(2\ell+1)\,  \hat{x}_{ijL}}{(\ell+1)(\ell+2)(2\ell+5) c^4}\partial_t^2 \Sigma_{ij}^{\rm EM} \Biggr\}, \\ 
    \label{eq: EM_current_multipole_moments}
    J_L^{\rm EM}=\, &{\rm FP}_{B=0}\varepsilon_{ab \langle i_\ell}\int d^3\mathbf{x}|\mathbf{x}|^B \Biggl\{\hat{x}_{L-1\rangle a}\Sigma_b^{\rm EM}\, -\frac{(2\ell+1)\hat{x}_{L-1\rangle ac}}{(\ell+2)(2\ell+3) c^2}\, \,\partial_t \Sigma_{bc}^{\rm EM}\Biggr\}.
\end{align}
The EM source densities entering these expressions, written in terms of the components of $A_\mu$ and the metric potentials of the parametrization \eqref{eq: ExpandedMetric}, are
\begin{subequations}
    \label{eq:app: EM_densities}
     \begin{align}
         \label{eq:app: EM_mass_density}
         \notag
         \Sigma^{\rm EM}&=\, -\frac{A_0 }{c^2}\Bigl(q_1 \delta_1 (\mathbf{x})+q_2 \delta_2 (\mathbf{x})\Bigr)\,+ \frac{1}{4 \pi c^4}\Biggl[2 A_i \partial_i \partial_j A_j\, +\partial_iA_i\partial_j A_j-\partial_i A_j \partial_jA_i\,-2\partial_t\Bigl( A_i \partial_iA_0\Bigr) \\ 
         &-8\pi VA_0\Bigl(q_1 \delta_1 (\mathbf{x})+q_2 \delta_2 (\mathbf{x})\Bigr)\, -2 A_0\partial_iV \partial_iA_0\, +4 V \partial_iA_0 \partial_iA_0\, -\frac{1}{2}\partial_t^2A_0^2\, \Biggr] + \mathcal{O}(1/c^6), \\ \notag
         \\
         \label{eq:app: EM_current_density}
         \Sigma_i^{\rm EM}&=\frac{1}{4\pi c^2}\Bigl[ \partial_i A_k \partial_k A_0\, -\partial_k A_i \partial_k A_0\Bigr]\,  + \mathcal{O}(1/c^4), \\ \notag
         \\
         \label{eq:app: EM_stress_density}
         \Sigma_{ij}^{\rm EM}&=\frac{1}{4\pi}\Biggl[\frac{1}{2}\delta_{ij}\partial_k A_0 \partial_k A_0\, -\partial_i A_0 \partial_jA_0\Biggr]\, +  + \mathcal{O}(1/c^2).
     \end{align}
\end{subequations}
%

%

The evaluation of the compact and non-compact terms follows the standard procedure described in Refs.~\cite{Blanchet:1995fg,Blanchet:2001aw}. We then specialize to the CoM frame and to circular orbits, using the reduction described in Sec.~\ref{subsec: CoM_circular_limit}. In particular, we impose $\mathbf{n}\cdot\mathbf{v}=0$ and $|\mathbf{v}|=\omega \, r$, with $\omega$ given in Eq.~\eqref{eq:orbital_frequency_circular_orbit}. The resulting source moments read

\begin{subequations}
\label{eq: source_multipole_moments_explicit}
    \begin{align}
    I_{ij}=& M\,\nu\,  \hat{r}_{ij}\,
+\frac{1}{c^2}\Biggl\{
\frac{11\,M\, \nu\, r^2}{21}(1-3\nu)\, \hat{v}_{ij}
-\frac{M^2\nu }{42 r}\left(1-\eta_1 \eta_2\right) (1+39\nu)\hat{r}_{ij}
\Biggr\} \\ \notag
&+\frac{1}{c^4}\Biggl\{
M^3 \hat{r}^{ij} \Biggl[
\frac{\nu }{1512 r^2}
\Bigl(
901\,\eta_1^2\eta_2^2
-3029\,\eta_1^2\eta_2^2\nu
+599\,\eta_1^2\eta_2^2\nu^2
\\ \notag
&\qquad
-486\,\eta_1^2
-6858\,\eta_1^2\nu
-486\,\eta_2^2
-6858\,\eta_2^2\nu
-461
-18395\nu
-241\nu^2
\Bigr)
\\ \notag
&\qquad
+\frac{\eta_1 \eta_2 \nu}{756 r^2}
\Bigl(
2912+938\nu-179\nu^2
\Bigr)
-\frac{\nu X_{12}}{28 r^2}
\Bigl(
9-97\nu
\Bigr)
(\eta_1^2-\eta_2^2)
\Biggr]
\\ \label{eq: Iij}
&\qquad
+M^2\hat{v}^{ij}
\Biggl[
\frac{1}{378}\eta_1\eta_2\nu
\Bigl(
79+415\nu-397\nu^2
\Bigr)r
+\frac{1}{378}\nu
\Bigl(
1607-1681\nu+229\nu^2
\Bigr)r
\Biggr]
\Biggr\},\\
       \label{eq: Iijk}
       I_{ijk}=&- M\, \nu\, X_{12} \Biggl\{\hat{r}_{ijk}\, +\frac{1}{c^2\, r}\Biggl[v_{\langle ij}r_{k\rangle}r^3(1-2\nu)\,-\hat{r}_{ijk}M\Bigl(\eta_1 \eta_2 (1-2\nu)-\nu \Bigr)  \Biggr] \Biggr\}, \\
       \label{eq: Iijkl}
       I_{ijkl}=&M\, \nu\Biggl\{ \hat{r}_{ijkl}(1-3\nu)\, +\frac{M}{100\, r\, c^2}\Biggl[\hat{r}_{ijkl}(3-125\nu+345\nu^2)\Bigl(1-\eta_1 \eta_2\Bigr)\, +156 v_{\langle ij}r_{k l\rangle} r^3(1-5\nu+5\nu^2) \Biggr]\Biggr\}, \\ 
       \label{eq: Jij}
       J_{ij}=&-M\nu\, X_{12}\, \varepsilon_{ab\langle j}\Biggl\{r_{i\rangle a}v_b\, +\frac{r_{i\rangle a}v_b\, M}{28 r\,c^2}\Biggl[ 67-8\nu-4\eta_1 \eta_2(1-2\nu)\Biggr] \Biggr\}, \\ 
  \notag
       J_{ijk}=&M \nu\varepsilon_{ab\langle k}\Biggl\{r_{ij\rangle a}v_b\, (1-3\nu)\, +\frac{1}{90 r c^2}\Biggl[14\,v_{ij\rangle b}r_a\, r^3 (1-5\nu+5\nu^2)\, \\      \label{eq: Jijk}
       &+M r_{ij\rangle a}v_b\Bigl(181-545\nu+65\nu^2-\eta_1 \eta_2(30-150\nu+30\nu^2)  \Bigr)\Biggr] \Biggr\}.
    \end{align}
\end{subequations}
\end{widetext}
All source multipole moments in Eq.~\eqref{eq: source_multipole_moments_explicit} are written in order-reduced form, namely with accelerations and their time derivatives replaced by using the lower-order equations of motion derived in Ref.~\cite{Placidi:2025xyi}.

We recall that the LO expressions for the moments $I_{ijklm}$, $I_{ijklmn}$, $J_{ijkl}$, and $J_{ijklm}$ do not depend on the charges and coincide with the standard GR results of Ref.~\cite{Blanchet:1995fg}. In the neutral limit, $\eta_1\to0$ and $\eta_2\to0$, all the source moments reduce to their known neutral-binary expressions~\cite{Blanchet:1995fg}.

The source multipole moments required to compute the instantaneous energy flux to 2PN order relative to the EM dipole are independently obtained also in the effective field theory approach. 
Specifically, $Q_i$ is evaluated to N${}^2$LO, $Q_{ij}$, $M_{i}$ and $I_{ij}$ are evaluated to NLO, $Q_{ijk}$, $M_{ij}$, $I_{ijk}$ and $J_{ij}$ are evaluated to leading order.
As a non-trivial validation of these results, the computations are extended to also include the terms required to explicitly check the conservation identities related to the Ward identities for the one-point function, corresponding to $\partial_\mu T^{\mu\nu} = 0$ and $\partial_\mu J_{\rm EM}^\mu = 0$. 

Including the additional terms required for the Ward identity checks, the calculation comprises 118 PN diagrams. After soft expansion, these diagrams involve massless two-point functions up to one loop order, which we evaluated by means of multi-loop techniques~\cite{Foffa:2011ub,Kol:2013ega,Foffa:2016rgu}.
The procedure used to generate and evaluate these diagrams has been implemented in a in-house computational framework developed in \texttt{Mathematica}. This framework interfaces with the \texttt{EFTofPNG} package~\cite{Levi:2017kzq} for PN expansion of the fundamental action, \texttt{xAct/xTensor} package~\cite{xAct} for tensor algebra, \texttt{FiniteFlow} package \cite{Peraro:2019svx} for finite field linear algebra evaluation, and the \texttt{LiteRed}, \texttt{Mint} and \texttt{Fermat} packages~\cite{Lee:2012cn,Lee:2013mka,Lee:2013hzt} for Feynman integral topology mapping and reduction through integration-by-parts identities.


\subsection{Tail contributions}
\label{subsec: tail_GW}
At the 2PN accuracy considered here, the tail contributions to the radiative moments are given by~\cite{Blanchet:1995fg}
\begin{subequations}
\begin{align}
&\delta U^{\rm tail}_{ij}
=
-\frac{2M}{c^3}
\int_0^{+\infty}d\tau
\left[
\ln\left(\frac{\tau}{2 b_0}\right)
+\frac{11}{12}
\right]
I_{ij}^{(4)}(T_R-\tau),
\\
&\delta V^{\rm tail}_{ij}
=
\frac{2M}{c^3}
\int_0^{+\infty}d\tau
\left[
\ln\left(\frac{\tau}{2 b_0}\right)
+\frac{7}{6}
\right]
J_{ij}^{(4)}(T_R-\tau),
\\
&\delta U^{\rm tail}_{ijk}
=
\frac{2M}{c^3}
\int_0^{+\infty}d\tau
\left[
\ln\left(\frac{\tau}{2 b_0}\right)
+\frac{97}{60}
\right]
I_{ijk}^{(5)}(T_R-\tau).
\end{align}
\end{subequations}
where $b_0$ is the arbitrary time scale entering the relation between harmonic and radiative coordinates, already introduced in Eq.~\eqref{eq:electric_dipole_tail_integral}. In the circular limit, these hereditary integrals can be evaluated explicitly, component by component, using the source moments derived in the previous subsection together with the integration formulas in Eq.~\eqref{eq:tail_int_formulas}.

We note that, at the accuracy required here, the source moments entering the tail integrals can always be truncated at LO, where they are independent of the charges. The charge-dependent corrections to the tail terms are therefore entirely due to the order-reduction procedure used to evaluate the time derivatives in the integrands.


\subsection{Memory contributions}
\label{subsec:memory_GW}

We finally consider the novel nonlinear memory contributions sourced by the EM field. In the MPM-PN construction, these terms arise from the EM part of the nonlinear source $\Lambda^{\mu\nu}$, Eq.~\eqref{eq: LambdaEM}, evaluated on the linear vector field
\begin{subequations}
\label{eq:linear_vector_field_QL_ML}
\begin{align}
&A_0^{\rm lin}
=
-\frac{1}{c^3}
\sum_{\ell\geq 0}
\frac{(-)^\ell}{\ell!}
\partial_L \tilde{Q}_L,
\\ \label{eq:linear_Ai}
&A_i^{\rm lin}
=
-\frac{1}{c^4}
\sum_{\ell\geq 1}
\frac{(-)^\ell}{\ell!}
\bigg[
\partial_{L-1}\dot{\tilde{Q}}_{iL-1}
\cr&\quad+
\frac{\ell}{\ell+1}
\varepsilon_{ijk}\partial_{jL-1}\tilde{M}_{kL-1}
\bigg],
\end{align}
\end{subequations}
where
\begin{equation}
\tilde{Q}_L\equiv \frac{Q_L(t-r/c)}{r},
\qquad
\tilde{M}_L\equiv \frac{M_L(t-r/c)}{r},
\end{equation}
and the dot in the first term of Eq.~\eqref{eq:linear_Ai} denotes a time derivative.
Physically, they correspond to gravitational radiation generated by the stress-energy tensor carried by the EM field. In addition to the genuine hereditary memory terms, there are local instantaneous contributions associated with the same quadratic interactions. 

At the accuracy required in this work, the relevant EM nonlinearities contribute to the radiative moments as follows:
\begin{subequations}
\label{eq:EM_nonlinear_interactions}
\begin{align}
\delta U_{ij}^{\rm mem\, EM}
&:\qquad
Q_i\times Q_j,
\quad
Q\times Q_{ij},
\\
\delta V_{ij}^{\rm mem\, EM}
&:\qquad
Q_i\times M_j,
\quad
Q\times M_{ij},
\\
\delta U_{ijk}^{\rm mem\, EM}
&:\qquad
Q_i\times Q_{jk},
\quad
Q\times Q_{ijk}.
\end{align}
\end{subequations}
Here $Q$ denotes the electric monopole, corresponding to the total charge of the source,
$Q=q_1+q_2$. The interactions involving two radiative EM multipoles give rise also to genuine hereditary memory terms,\footnote{The hereditary term associated with the $Q_i \times Q_i$ interaction is the EM analogue of the dipolar memory obtained in scalar-tensor theory \cite{Bernard:2022noq}.} while the interactions involving $Q$ are Coulomb-radiative nonlinearities and contribute through local instantaneous terms. 

The explicit expressions are obtained by performing the MPM iteration at linear order in $1/R$, applying the TT projection, and matching the resulting waveform to the general radiative expansion in Eq.~\eqref{eq:TT_projection}. We do not reproduce the intermediate MPM-PN iteration here, and refer any reader interested in these details to the standard matched derivations of nonlinear memory, see e.g. Refs.~\cite{Blanchet:1997ji,Blanchet:1997jj}.

The resulting charge-dependent nonlinear contributions to the radiative moments in radiative coordinates are given, in terms of the source moments, by
\begin{widetext}
\begin{subequations}
\label{eq:EM_memory_results}
\begin{align}
\delta U_{ij}^{\rm mem\, EM}
=&
-\frac{1}{3c^3}
\left[\int_{0}^{+\infty} d\tau\, Q_{\langle i}^{(2)}Q_{j\rangle}^{(2)}(T_R-\tau)+
2 Q_{\langle i}^{(1)} Q_{j\rangle}^{(2)}
+
2 Q_{\langle i} Q_{j\rangle}^{(3)}
\right]
-\frac{Q}{3c^3} Q_{ij}^{(3)},
\\
\delta V_{ij}^{\rm mem\, EM}
=&
\frac{1}{8c^3}
\left[
3 Q_{\langle i}^{(1)} M_{j\rangle}^{(2)}
-3 M_{\langle i}^{(1)} Q_{j\rangle}^{(2)}
+Q_{\langle i} M_{j\rangle}^{(3)}
-M_{\langle i} Q_{j\rangle}^{(3)}
\right]
+\frac{Q}{6c^3} M_{ij}^{(3)},
\\
\delta U_{ijk}^{\rm mem\, EM}
=&
-\frac{1}{c^3}
\Bigg[\frac{3}{5}\int_{0}^{+\infty} d \tau\, Q_{\langle k}^{(2)}(t) Q_{ij\rangle}^{(3)}(T_R-\tau)+
\frac{6}{5} Q_{\langle k}^{(2)} Q_{ij\rangle}^{(2)}
+\frac{9}{5} Q_{\langle ij}^{(1)} Q_{k\rangle}^{(3)}
+\frac{1}{5} Q_{\langle k}^{(1)} Q_{ij\rangle}^{(3)}
\nonumber\\
&\hspace{3.0cm}
+\frac{7}{10} Q_{\langle ij} Q_{k\rangle}^{(4)}
+\frac{3}{10} Q_{\langle k} Q_{ij\rangle}^{(4)}
\Bigg]
-\frac{7Q}{30c^3} Q_{ijk}^{(4)} .
\end{align}
\end{subequations}
\end{widetext}
The time integrals are the genuine hereditary memory pieces mentioned above, while the remaining local terms are the associated instantaneous completions required by the extraction of the radiative moments. The last term in each line comes from the Coulomb-radiative interaction between the electric monopole and the corresponding radiative EM multipole.

The hereditary integrals in Eq.~\eqref{eq:EM_memory_results} are evaluated in the circular limit by decomposing the EM source moments into harmonics of the orbital frequency, in the same way as for the tail integrals. It should be noted, however, that this procedure applies only to the oscillatory contributions. The zero-frequency pieces, which give rise to the so-called direct-current (DC) memory, are not included in the present analysis and will be addressed in future work.
A complete treatment of the DC memory would also require including the charge-dependent corrections to the standard nonlinear memory already present for neutral binaries. Although these effects are formally of higher PN order, starting at relative 2.5PN order, their zero-frequency character leads to a secular accumulation over the radiation-reaction timescale. As a result, they can contribute at LO and must be treated consistently together with the EM-sourced DC terms.
\section{Spherical-harmonic waveform modes}
\label{sec:waveformModes}

Having determined the charge-corrected radiative STF multipole moments of the gravitational field at infinity, we now derive the corresponding spherical-harmonic modes. This form is particularly useful for waveform modeling, since the two gravitational-wave polarizations are naturally decomposed on the basis of spin-weighted spherical harmonics as
\begin{equation}
    h_+ - i h_\times
    =
    \sum_{\ell\geq 2}\sum_{m=-\ell}^{\ell}
    h_{\ell m}\,{}_{-2}Y_{\ell m}(\Theta,\Phi).
\end{equation}
The complex modes $h_{\ell m}$ provide a mode-by-mode description of the radiation field, directly connected to the quantities used in analytical waveform models and numerical-relativity comparisons. They also allow the energy flux to be written as a sum over individual angular modes. In this section, we project the radiative STF moments derived above onto this spherical basis and compute the charge-dependent corrections to the modes needed at 2PN accuracy.

The harmonic modes $h_{\ell m}$ are defined in terms of the mass-type $(U_{\ell m})$ and the current-type $(V_{\ell m})$ radiative multipoles
\begin{equation}
    \label{eq: generic_modes}
h^{\ell m}=\, -\frac{G}{\sqrt{2}\, R\,  c^{\ell+2}}\left[U^{\ell m}\, -\frac{i}{c}V^{\ell m} \right],
\end{equation}
where as before $R$ is the distance of the source in the radiative coordinates. We note that, for non-precessing binaries, the mode $h^{\ell m}$ can be determined by the mass-type multipoles in the case of $\ell+m$ even, and by the current-type multipoles when $\ell+m$ is odd~\cite{Faye:2012we}, i.e.
\begin{subequations}
    \begin{align}
        \label{eq: hlm_even}
        h^{\ell m}=&-\frac{ U^{\ell m}}{\sqrt{2}R\, c^{\ell+2}}, \quad \text{if $\ell+m$ is even}, \\
         \label{eq: hlm_odd}
        h^{\ell m}=&\frac{ V^{\ell m}}{\sqrt{2}R\, c^{\ell+3}} , \quad \text{if $\ell+m$ is odd}. 
    \end{align}
\end{subequations}

The radiative spherical multipole moments are determined from the STF radiative moments $U_L$ and $V_L$, computed in Sec.~\ref{sec:MassCurrentMultipoles}, through
\begin{subequations}
\begin{align}
\label{eq: Ulm}
U^{\ell m}
&=
\frac{4}{\ell!}
\sqrt{\frac{(\ell+1)(\ell+2)}{2\ell(\ell-1)}}
\,\alpha_L^{\ell m}U_L,
\\
\label{eq: Vlm}
V^{\ell m}
&=
-\frac{8}{\ell!}
\sqrt{\frac{\ell(\ell+2)}{2(\ell+1)(\ell-1)}}
\,\alpha_L^{\ell m}V_L.
\end{align}
\end{subequations}
The STF tensors $\alpha_L^{\ell m}$ relate the STF products $\hat N_L$ to the ordinary scalar spherical harmonics according to
\begin{equation}
\label{eq:alpha_STF_definition}
\hat N_L(\Theta,\Phi)
=
\sum_{m=-\ell}^{\ell}
\alpha_L^{\ell m}Y^{\ell m}(\Theta,\Phi),
\end{equation}
or, equivalently,
\begin{equation}
\alpha_L^{\ell m}
=
\int d\Omega\,\hat N_L(\Theta,\Phi)\,\overline{Y}^{\ell m}(\Theta,\Phi),
\end{equation}
where the overbar denotes the complex conjugation.

As a representative result, and in view of its phenomenological relevance, we display below the circular-orbit expression of the dominant $(\ell,m)=(2,2)$ waveform mode. The circular limit is taken as described in Sec.~\ref{subsec: CoM_circular_limit}.
Using harmonic coordinates and including all charge-dependent corrections required at 2PN accuracy, we obtain
\begin{widetext}
\begin{equation}
\label{eq: h22_total}
     h^{22}=\,8\sqrt{\frac{\pi}{5}} e^{-2 i \phi}\, M\, \nu\, x\left\{\left(1-\eta_1 \eta_2\right)^{2/3}\,+x\, H^{22}_{\rm inst,NLO}\,+x^{3/2}\, \left(H^{22}_{\rm tail, LO}\, +H^{22}_{\rm mem, LO}\right)\,+x^2\, H^{22}_{\rm inst,N^2LO}\right\},
\end{equation}
where
\begin{subequations}
  \begin{align}
  \label{eq: H22_1PN}
  \notag
   &H^{22}_{\rm inst, NLO}=\,-\frac{1}{42\, (1-\eta_1\eta_2)^{2/3}}\Bigl[107-55 \nu +14 \eta_2^2+14 \eta_1^2+\eta_1^2 \eta_2^2 (37-55 \nu )\\ 
   &\hspace{2cm} -2 \eta_1 \eta_2 (86-55 \nu )\,+14\, X_{12} \left( \eta_1^2-\eta_2^2\right) \Bigr], \\ \label{eq: H22_tail}
   &H^{22}_{\rm tail, LO}=\,2 (1-\eta_1 \eta_2)^{2/3} \left(\pi +3 i \log
   \left(\frac{x}{x_0}\right)\right),\\ \label{eq: H22_memory}
   &H^{22}_{\rm mem, LO}=\,\frac{i \nu}{24}   (\eta_1-\eta_2)^2 (1-\eta_1 \eta_2)^{2/3}, \\
   \notag
    &H^{22}_{\rm inst, N^2LO}=\,-\frac{1}{1512\,\left(1-\eta_1 \eta_2\right)^{2}}\Biggl[ 2173+7483 \nu -2047 \nu ^2+2 \eta_1^3 \eta_2 \left(174-8208 \nu
   -\eta_2^2 (1877-1648 \nu -4346 \nu ^2)\right)\\ \notag
    &+252 \eta_2^4 (1-2 \nu )-276
   \eta_2^2 (2-27 \nu )-2 \eta_1 \eta_2
   \left(5549+1796 \nu -4178 \nu ^2-6 \eta_2^2 (29-1368 \nu
   )\right)\\ \notag
    &+6 \eta_1^2 \left(46 (27 \nu-2 )-2 \eta_2^4
   (25-747 \nu )+\eta_2^2 (2489-1451 \nu -2131 \nu
   ^2)\right)\\ \notag
    &-\eta_1^4 \left(\eta_2^2 (300-8964 \nu
   )+252 (-1+2 \nu )+\eta_2^4 (1751-2527 \nu +2215 \nu
   ^2)\right)
   \\ \label{eq:H22_2PN}
    &-12\left(\eta_1^2-\eta_2^2\right)\Bigl(46-21 \eta_2^2+261 \nu -29 \eta_1\eta_2 (1+18
   \nu )+ -21\eta_1^2+\eta_1^2\eta_2^2
   (25+261 \nu )\Bigr)\Biggr].
\end{align}  

\end{subequations}
\end{widetext}

Analogous expressions for all the other waveform modes entering at 2PN accuracy are collected in App.~\ref{app: waveform_modes}.

\section{Energy flux}
\label{sec:EnergyFlux}

The 2PN-accurate energy flux emitted by a charged binary can be split into two contributions,
\begin{equation}
    \label{eq: total_flux}
    \mathcal{F}_{\rm total}= \mathcal{F}_{\rm V}+\mathcal{F}_{\rm T},
\end{equation}
where $\mathcal{F}_{\rm V}$ denotes the vector-field contribution and $\mathcal{F}_{\rm T}$ the tensor-field contribution. The vector flux starts at order $1/c^3$, while the tensor flux starts, as usual, at order $1/c^5$. 

In the following subsections we derive the two contributions to the flux in terms of the radiative multipole moments computed in the previous sections. We then illustrate the impact of the new charge-dependent corrections through representative plots for selected charge configurations. Although a 2PN-accurate total flux, counted relative to the leading vector emission, would require the tensor flux only through NLO, we also compute its N${}^2$LO contribution because of its relevance for waveform modeling and for comparison with the known neutral result.

Furthermore, the instantaneous energy flux to 2PN  order has been independently derived within the effective field theory approach, finding full agreement.
The hereditary terms are evaluated within the PN-matched MPM framework.

We specify that all results of this section are expressed in order-reduced form, namely with all accelerations and their time derivatives eliminated using the lower-order equations of motion~\cite{Placidi:2025xyi}.
\subsection{Vector flux}
\label{subsec:VectorFlux}

The vector flux has a leading dipolar contribution and therefore starts at order $1/c^3$, one PN order earlier than the leading quadrupolar contribution to the tensor flux. As shown in Ref.~\cite{Placidi:2025xyi}, this dipolar emission affects the dynamics already at 1.5PN order, generating dissipative terms in the equations of motion.

The vector flux is written in terms of the radiative electric and magnetic multipole moments derived in Sec.~\ref{sec:ElectricMagneticMultipoles}. It is given by~\cite{Khalil:2018aaj,Henry:2023len}
\begin{align}
& \mathcal{F}_{\rm V}
=
\sum_{\ell\geq 1}
\frac{1}{c^{2\ell+1}\ell!(2\ell+1)!!}
\Bigg[
\frac{\ell+1}{\ell}
\mathcal{E}_L^{(1)}
\mathcal{E}_L^{(1)}
\cr&\qquad+
\frac{\ell}{c^2(\ell+1)}
\mathcal{B}_L^{(1)}
\mathcal{B}_L^{(1)}
\Bigg].
\end{align}
Inserting the relation between radiative and source moments discussed in Sec.~\ref{sec:ElectricMagneticMultipoles}, the vector flux can be decomposed as
$\mathcal{F}_{\rm V}=\mathcal{F}_{\rm V}^{\rm inst}+\mathcal{F}_{\rm V}^{\rm tail}$.
In terms of the source moments, the instantaneous contribution reads
\begin{align}
      &\mathcal{F}_{\rm V}^{\rm inst}=\frac{2}{3c^3}Q_i^{(2)}Q_i^{(2)}\, +\frac{1}{c^5}\Biggl[\frac{1}{6}M_i^{(2)}M_i^{(2)}\, +\frac{1}{20}Q_{ij}^{(3)}Q_{ij}^{(3)}\Biggr] \cr 
    &\quad+\frac{1}{c^7}\Biggl[\frac{1}{45}M_{ij}^{(3)}M_{ij}^{(3)}\,+\frac{2}{945}Q_{ijk}^{(4)}Q_{ijk}^{(4)}\Biggr] +\mathcal{O}(1/c^9),
\end{align}
while the leading hereditary contribution is the electric-dipole tail,
\begin{equation}
    \mathcal{F}_{\rm V}^{\rm tail}=\frac{2 M}{ c^6}Q_i^{(2)}(T_R)\int_0^{\infty} d\tau\,  \bigg[
\ln\left(\frac{\tau}{2b_0}\right)+\frac{3}{4}
\bigg]Q_i^{(4)}(T_R-\tau).
\end{equation}

Using the explicit expressions for the electric and magnetic multipole moments in Eqs.~\eqref{eq: electric_magnetic_multipoles_moments_explicit}, and rewriting the result in the CoM frame and for circular orbits, in terms of the gauge-invariant parameter $x$, the 2PN-accurate vector flux takes the schematic form
\begin{align}
\notag
&\mathcal{F}_{\rm V}
=\,
\frac{2x^4\nu^2}{3}
\Biggl\{
\left(1-\eta_1\eta_2\right)^{2/3}
\left(\eta_1-\eta_2\right)^2
\\
&\quad
+x\,f_{\rm V}^{\rm inst,NLO}
+x^{3/2}f_{\rm V}^{\rm tail,LO}
+x^2f_{\rm V}^{\rm inst,N^2LO}
\Biggr\}.
\label{eq: Vector_flux_explicit_expression}
\end{align}
The instantaneous coefficients $f_{\rm V}^{\rm inst,NLO}$ and $f_{\rm V}^{\rm inst,N^2LO}$ are rather lengthy and are provided explicitly in App.~\ref{app:explicit_results_energy_flux}. The tail coefficient, instead, is sufficiently compact and reads
\begin{equation}
    f_V^{\rm tail,LO}=\, 2\pi \bigl(1-\eta_1 \eta_2\bigr)^{2/3} \bigl(\eta_1-\eta_2\bigr)^2.
\end{equation}

The complete instantaneous vector flux through relative 2PN order is independently computed from the source multipole moments obtained within the effective field theory approach, finding exact agreement with the instantaneous part of Eq.~\eqref{eq: Vector_flux_explicit_expression} reported in App.~\ref{app:explicit_results_energy_flux}.

We checked that the total vector flux computed here agrees with the 1PN result of Ref.~\cite{Julie:2018lfp}. Moreover, as expected, it vanishes in the neutral limit, namely for $\eta_1\to0$ and $\eta_2\to0$.

\subsection{Tensor flux}
\label{subsec:TensorFlux}

As for neutral binaries, the tensor flux has a leading quadrupolar contribution and therefore starts at order $1/c^5$. In terms of the TT waveform, it is defined by~\cite{Blanchet:1995fg,Blanchet:1997jj}
\begin{align}
\notag
\mathcal{F}_{\rm T}
&=
\frac{c^3R^2}{32\pi}
\int d\Omega\,
\left(
\frac{\partial h_{ij}^{\rm TT}}{\partial T_R}
\right)
\left(
\frac{\partial h_{ij}^{\rm TT}}{\partial T_R}
\right)
\\
\notag
&=
\sum_{\ell=2}^{+\infty}
\frac{1}{c^{2\ell+1}}
\Biggl[
\frac{(\ell+1)(\ell+2)}
{(\ell-1)\ell\,\ell!(2\ell+1)!!}
U_L^{(1)}U_L^{(1)}
\\
&\qquad
+
\frac{4\ell(\ell+2)}
{c^2(\ell-1)(\ell+1)!(2\ell+1)!!}
V_L^{(1)}V_L^{(1)}
\Biggr].
\label{eq: tensor_flux_definition}
\end{align}
Substituting the relations between radiative and source moments discussed in Sec.~\ref{sec:MassCurrentMultipoles}, the tensor flux can be decomposed, similarly to the vector flux, as
$\mathcal{F}_{\rm T}=\mathcal{F}_{\rm T}^{\rm inst}+\mathcal{F}_{\rm T}^{\rm tail}$.
The instantaneous contribution reads
\begin{align}
\notag
\mathcal{F}_{\rm T}^{\rm inst}
=&\,
\frac{1}{5c^5}I_{ij}^{(3)}I_{ij}^{(3)}
+\frac{1}{c^7}
\left[
\frac{1}{189}I_{ijk}^{(4)}I_{ijk}^{(4)}
+\frac{16}{45}J_{ij}^{(3)}J_{ij}^{(3)}
\right]
\\
&+
\frac{1}{c^9}
\left[
\frac{1}{9072}I_{ijkl}^{(5)}I_{ijkl}^{(5)}
+\frac{1}{84}J_{ijk}^{(4)}J_{ijk}^{(4)}
\right]
+\mathcal{O}(1/c^{11}),
\label{eq: tensor_flux_instantaneous_part}
\end{align}
Substituting the relations between radiative and source moments discussed in Sec.~\ref{sec:MassCurrentMultipoles}, the tensor flux can be decomposed, in analogy with the vector flux, as
$\mathcal{F}_{\rm T}=\mathcal{F}_{\rm T}^{\rm inst}+\mathcal{F}_{\rm T}^{\rm tail}$.
The instantaneous contribution reads
\begin{align}
\notag
\mathcal{F}_{\rm T}^{\rm inst}
=&\,
\frac{1}{5c^5}I_{ij}^{(3)}I_{ij}^{(3)}
+\frac{1}{c^7}
\left[
\frac{16}{45}J_{ij}^{(3)}J_{ij}^{(3)}+\frac{1}{189}I_{ijk}^{(4)}I_{ijk}^{(4)}
\right]
\\
&+
\frac{1}{c^9}
\left[
\frac{1}{84}J_{ijk}^{(4)}J_{ijk}^{(4)}+\frac{1}{9072}I_{ijkl}^{(5)}I_{ijkl}^{(5)}
\right]
+\mathcal{O}(1/c^{11}),
\label{eq: tensor_flux_instantaneous_part}
\end{align}
while the hereditary sector is given, at this accuracy, by the mass-quadrupole tail~\cite{Blanchet:1995fg},
\begin{align}
&\mathcal{F}_{\rm T}^{\rm tail}
=
\frac{4M}{5c^8}
I_{ij}^{(3)}(T_R)
\int_0^{+\infty}d\tau\,
\bigg[
\ln\left(\frac{\tau}{2b_0}\right)
\cr&\qquad+\frac{11}{12}
\bigg]
I_{ij}^{(5)}(T_R-\tau).
\label{eq:tensor_flux_tail}
\end{align}
We have explicitly checked that the new charge-dependent oscillatory-memory contributions to the radiative moments do not contribute to the energy flux at this order. This cancellation is consistent with the harmonic structure of the circular problem: the oscillatory memory terms are either orthogonal to the leading instantaneous contribution in the STF contractions entering the flux, or contribute only through terms quadratic in the memory amplitude, which are beyond the accuracy considered here.

Using the explicit expressions for the mass and current source moments obtained in Sec.~\ref{sec:MassCurrentMultipoles}, the tensor flux in the CoM frame and for circular orbits takes the form
\begin{align}
\notag
\mathcal{F}_{\rm T}
=&\,
\frac{32x^5\nu^2}{5}
\Biggl\{
\left(1-\eta_1\eta_2\right)^{4/3}
+x f_{\rm T}^{\rm inst,NLO}
\\
&\qquad
+x^{3/2}f_{\rm T}^{\rm tail,LO}
+x^2 f_{\rm T}^{\rm inst,N^2LO}
\Biggr\}.
\label{eq:Tensor_Energy_flux}
\end{align}
The instantaneous coefficients $f_{\rm T}^{\rm inst,NLO}$ and
$f_{\rm T}^{\rm inst,N^2LO}$ are provided explicitly in
App.~\ref{app:explicit_results_energy_flux}, while the tail coefficient is
\begin{equation}
f_{\rm T}^{\rm tail,LO}
=
4\pi\left(1-\eta_1\eta_2\right)^{4/3}.
\end{equation}

The instantaneous tensor flux through NLO, is independently obtained also within effective field theory approach, and agrees exactly with the corresponding terms of Eq.~\eqref{eq:Tensor_Energy_flux} reported in App.~\ref{app:explicit_results_energy_flux}.

In the neutral limit, $\eta_1\to0$ and $\eta_2\to0$, the tensor flux reduces to the well-known gravitational-wave energy flux emitted by a neutral binary system~\cite{Blanchet:1995fg}. At the 1PN level, it also agrees with the charged-binary result of Ref.~\cite{Khalil:2018aaj}.


\subsection{Impact of charge on the energy flux}
\label{sec:Charge_impact_flux}

Finally, we investigate numerically how the electric charges affect the energy flux emitted by a charged black-hole binary.

We begin by evaluating the tensor energy flux and comparing it with the corresponding neutral case. In Fig.~\ref{fig:tensor_flux_uguali}, we plot the tensor flux~\eqref{eq:Tensor_Energy_flux} as a function of the PN parameter $x$~\eqref{eq: x_definition} for fixed charge magnitudes. The left panel corresponds to equal-sign charges, $\eta_1=\eta_2=0.8$, while the right panel shows the opposite-sign configuration, $\eta_1=0.8$ and $\eta_2=-0.8$. The results show that, as expected, the EM interaction produces appreciable deviations from the neutral case when charge magnitudes are sufficiently large. These plots also allow us to assess how the tensor energy flux changes as the PN accuracy is increased, and to compare the effect of the electric charges with that of higher-order PN corrections.

By comparing the two panels, we can also investigate how the nature of the EM interaction affects the emitted flux. When the two charges have the same sign, corresponding to a repulsive EM interaction, the tensor energy flux is systematically reduced with respect to the neutral case. In contrast, for opposite-sign charges, corresponding to an attractive EM interaction, the emitted energy flux is enhanced relative to the neutral configuration.

The effect becomes even more pronounced when one of the charged black holes reaches extremality.\footnote{We do not consider the simultaneous same-sign extremal limit, $\eta_1=\eta_2=1$. In this case the leading gravitational attraction is exactly balanced by the Coulomb repulsion, so that the circular-orbit condition degenerates, $\omega^2\propto(1-\eta_1\eta_2)\to0$. Correspondingly, both the binding energy and the total radiated flux vanish in our circular-orbit expressions. This limit should therefore be interpreted as a no-force configuration, rather than as an ordinary quasi-circular inspiral within the PN expansion used here.} This is illustrated in Fig.~\ref{fig:tensor_flux_estremale1}, where we consider the same charge configurations as in Fig.~\ref{fig:tensor_flux_uguali}, but with $\eta_1=1$. In this case, the influence of the electric charges on the tensor energy flux is significantly enhanced, leading to larger deviations from the neutral case. In both configurations, however, the differences remain negligible in the small-$x$ regime, where all curves nearly overlap.

We then turn to the total flux. In Fig.~\ref{fig:total_flux}, we consider the N${}^2$LO total flux $\mathcal{F}$~\eqref{eq: total_flux}, namely through $\mathcal{O}(c^{-7})$ relative to the leading vector contribution. At this accuracy, the vector flux is included up to N${}^2$LO, while the tensor flux enters, at most, at NLO.

To study the total flux for fixed values of the charges, we employ a Padé resummation of the PN expansion, choosing a diagonal Padé on the variable $y\equiv \sqrt{x}$. The advantage of this resummation is clearly illustrated in the left panel of Fig.~\ref{fig:total_flux}, where the Padé-resummed total flux is compared with its straightforward Taylor expansion. The two predictions differ significantly. More importantly, the Taylor-expanded flux becomes negative beyond a certain value of $x$, leading to an unphysical behavior, since the emitted energy flux must remain positive. This pathology originates from the oscillatory behavior of the truncated PN series and is effectively cured by the Padé resummation, which incorporates the available PN information into a more robust approximation. As shown in the figure, the Padé-resummed flux remains well behaved throughout the entire range of $x$ considered.

It is worth emphasizing that these oscillations affect the total flux, but were not observed in the tensor flux shown in the previous figures. The reason is that those figures display the tensor flux computed through N${}^2$LO, where the hereditary tail contributions are already present and help stabilize the PN expansion. In contrast, for the total flux considered here, the tensor contribution enters only at NLO, while the corresponding tail terms appear at higher PN order and therefore cannot mitigate the oscillatory behavior. 

\begin{widetext}

 \begin{figure}[H]
    \centering
\includegraphics[scale=0.65]{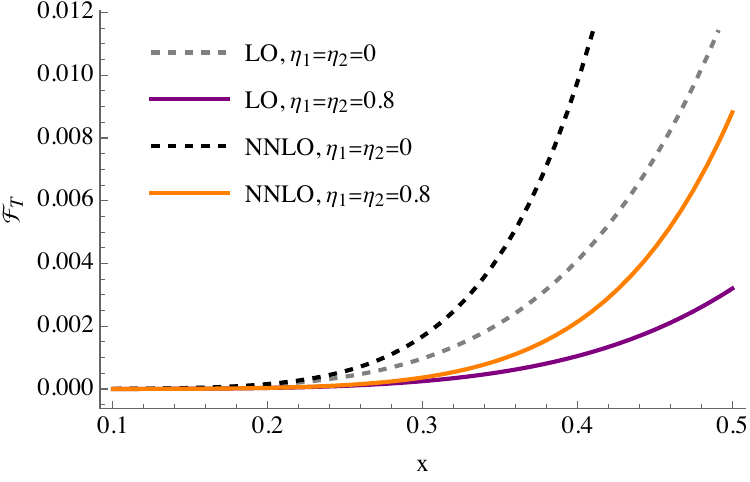} 
\includegraphics[scale=0.65]{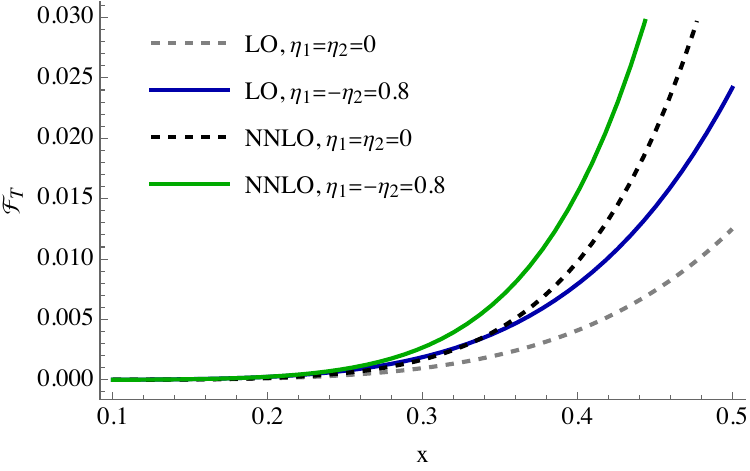}
    \caption{Tensor energy flux as a function of the parameter $x$ for equal-magnitude charges. The left panel corresponds to charges of the same sign, whereas the right panel corresponds to oppositely charged black holes. The dashed curves denote the tensor energy flux in the neutral limit, at Newtonian order (gray) and at N${}^2$LO order (black). The solid curves show the tensor energy flux in the charged case, at Newtonian order (purple in the left, blue in the right) and at N${}^2$LO order (orange in the left, green in the right).}
    \label{fig:tensor_flux_uguali}
\end{figure}
 \begin{figure}[H]
    \centering
\includegraphics[scale=0.65]{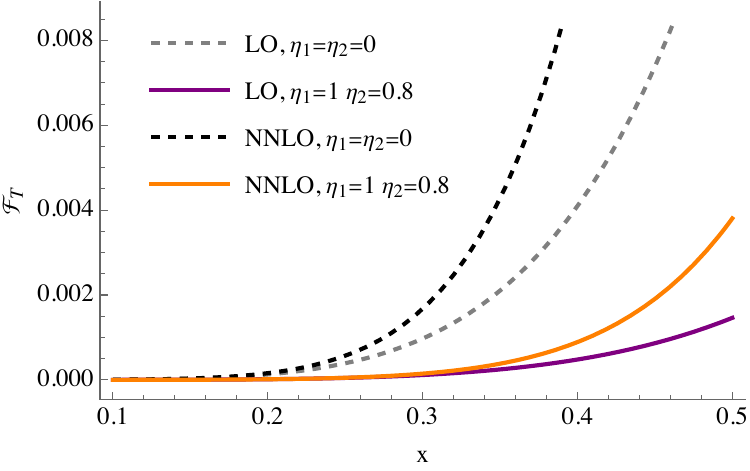} 
\includegraphics[scale=0.65]{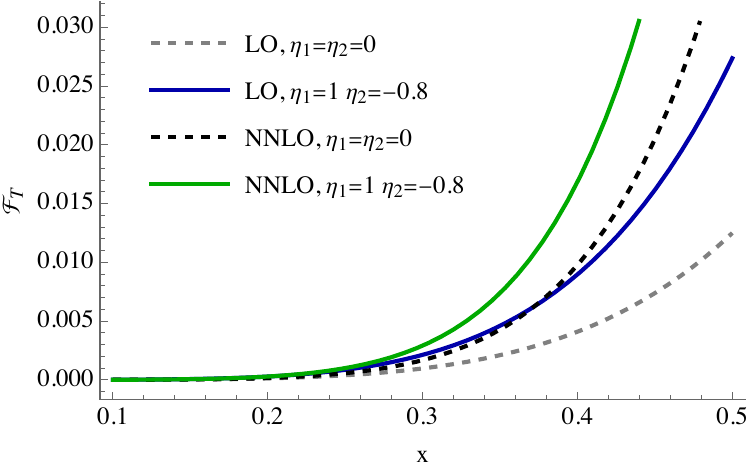}
    \caption{Tensor energy flux as a function of the parameter $x$ for the case in which one of two black holes is extremal. The left panel corresponds to charges of the same sign, whereas the right panel corresponds to oppositely charged black holes. The dashed curves denote the tensor energy flux in the neutral limit, at Newtonian order (gray) and at N${}^2$LO order (black). The solid curves show the tensor energy flux in the charged case, at Newtonian order (purple in the left, blue in the right) and at N${}^2$LO order (orange in the left, green in the right).}
    \label{fig:tensor_flux_estremale1}
\end{figure}
 \begin{figure}[H]
    \centering
\includegraphics[scale=0.65]{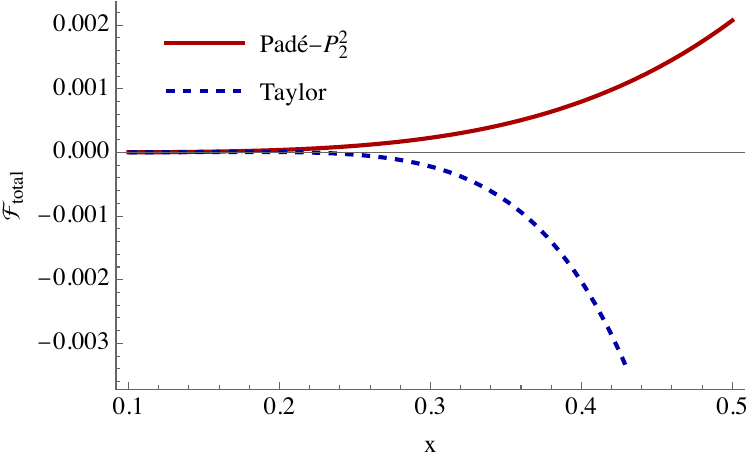} 
\includegraphics[scale=0.65]{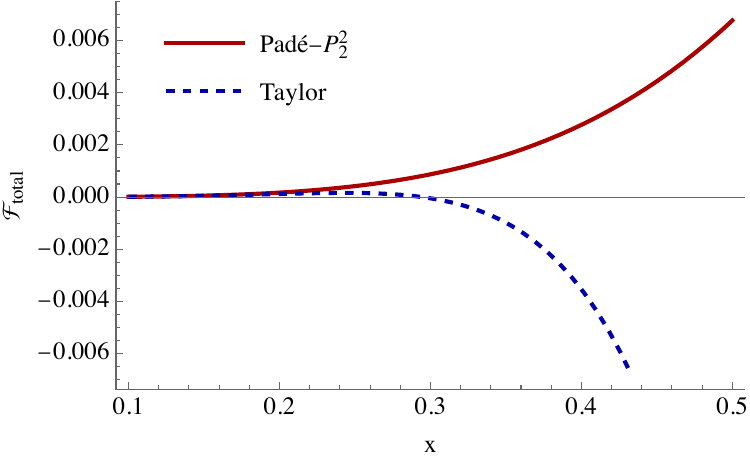}
    \caption{ Total flux at the N${}^2$LO order as a function of the parameter $x$. In the left panel, $\eta_1 = 0.8$ and $\eta_2 = 0.5$, while in the right panel, $\eta_1 = 0.8$ and $\eta_2 = -0.5
    $. The solid red curve represents the total flux after Padé resummation, whereas the dashed blue curve corresponds to the Taylor-expanded result.}
    \label{fig:total_flux}
\end{figure}
\end{widetext}

\section{Conclusion}
\label{sec:conclusion}

In this paper, we analyzed the radiation emitted by a binary system of electrically charged black holes. Building on the 2PN dynamics derived in Ref.~\cite{Placidi:2025xyi}, we extended the analysis to the radiation sector and computed the charge-dependent waveform modes and energy flux for quasi-circular binaries.
The source multipoles entering the instantaneous energy flux were computed independently within two complementary formulations of the PN expansion, the MPM-PN and the effective field theory formalisms.

We first solved the field equations for the gravitational and EM sectors, taking into account the nonlinear interactions between the two fields. From the PN-expanded field equations, we derived the effective source densities needed to compute the source and radiative multipole moments at 2PN accuracy. In parallel, the effective field theory calculation computes the one-point radiation function employing the background field formulation, yielding independent predictions for the N${}^2$LO electromagnetic and NLO tensor source multipoles, further checking explicitly the Ward identities for the one-point radiation function.

For the vector field, we computed the electric and magnetic source moments $Q_L$ and $M_L$ required for the energy flux. The corresponding radiation is dominated by the leading electric-dipole contribution, which is absent in the neutral case. Within the MPM-PN matched formalism, these source moments are related to the radiative moments $\mathcal{E}_L$ and $\mathcal{B}_L$. At the accuracy considered here, the only nonlinear propagation effect in the vector sector is the leading tail correction to the radiative electric dipole, entering at relative 1.5PN order. As expected, all vector multipole moments are proportional to the charges and vanish in the neutral limit. At NLO, our results agree with those of Ref.~\cite{Julie:2018lfp}.

We then computed the mass- and current-type source moments $I_L$ and $J_L$ entering the gravitational radiation field. Particular attention was devoted to the new charge-dependent contributions induced by the EM field, both through the EM part of the stress-energy tensor and through the nonlinear interaction terms in $\Lambda^{\mu\nu}$. Additional charge-dependent corrections arise from the order reduction procedure and from the transformation to the CoM frame, which must be performed using the charged EoMs. The source moments were then related to the radiative moments $U_L$ and $V_L$ through the MPM-PN matched formalism. Besides the usual tail terms, whose explicit expressions acquire charge dependence through the source moments and the orbital dynamics, we found new nonlinear contributions sourced by the EM stress-energy tensor. These include genuine oscillatory-memory terms, generated by products of radiative EM multipoles, as well as Coulomb-radiative nonlinearities involving the electric monopole.

Using the radiative multipole moments, we derived the spin-weighted spherical-harmonic waveform modes and the total energy flux. The latter naturally splits into a vector contribution, starting with the leading electric-dipole flux, and a tensor contribution, which reduces to the standard gravitational-wave flux in the neutral limit.
The instantaneous vector flux to N${}^2$LO and the instantaneous tensor flux to NLO were independently computed also within the effective field theory approach, with available results in exact agreement between the two formalisms.
We further checked that the vector flux agrees with the 1PN result of Ref.~\cite{Julie:2018lfp} and that the tensor flux reproduces the known neutral-binary result when the charges are set to zero. We also studied the impact of charge on the flux for representative charge configurations, including cases in which one of the black holes is extremal. These examples show that charge can produce significant deviations from the neutral flux, with the sign of the EM interaction determining whether the tensor contribution is enhanced or suppressed.

The analysis presented here was restricted to quasi-circular orbits. A natural extension will be the generalization to eccentric binaries, where the interplay between dipolar EM radiation, hereditary effects, and orbital harmonics is expected to be richer. In addition, the present work did not include the zero-frequency, or DC, memory contributions. Their consistent treatment, together with the associated charge-dependent corrections to the standard nonlinear memory, is left for future work.

\acknowledgments
M.P. wishes to thank Nicola Bartolo, Pierpaolo Mastrolia and Angelo Ricciardone for useful and insightful discussions.
A.P., E.G., and  M.O. acknowledge financial support from
the Italian Ministry of University and Research (MUR)
through the program “Dipartimenti di Eccellenza 2018-
2022” (Grant SUPER-C). M. O. acknowledges support from “Fondo di Ricerca d’Ateneo”
2023 (GraMB) of the University of Perugia. E.G, and M.O. acknowledge support by the “Center of Gravity”, which is
a Center of Excellence funded by the Danish National
Research Foundation under grant No. 184. 
M.P.’s research is supported by the European Union under the Next Generation EU programme.
M.P. acknowledges the support of the INFN initiatives \textit{Amplitudes} and \textit{InDark}.
\appendix
\section{Explicit coefficients energy flux}
\label{app:explicit_results_energy_flux}
\begin{widetext}
The explicit expressions of the PN-coefficients in Eqs.~\eqref{eq: Vector_flux_explicit_expression} are
\begin{subequations}
\begin{align} 
\notag
  f_V^{\rm inst,NLO}=&\frac{1}{15 \left(1-\eta_1 \eta_2\right)^{2/3}}\Biggl[5 X_{12} \left(8 \eta_1^3 \eta_2-8 \eta_1
   \eta_2^3+\eta_2^2 (3+\eta_2^2)-\eta_1^4 (1+3 \eta_2^2)-3
   \eta_1^2 (1-\eta_2^4)\right)\\ \notag
   &-5 \eta_2^2 \left(9+\eta_2^2+4 \nu \right)-5 \eta_1^4
   \left(1+\eta_2^2 (-2+4 \nu )\right)+\eta_1^3
   \eta_2 \left(55+40 \nu +8 \eta_2^2 (2+5 \nu
   )\right)\\ \label{eq: Vector_flux_explicit_expression_NLO_coefficient}
   &+\eta_1 \eta_2 \left(126+40 \nu +5
   \eta_2^2 (11+8 \nu )\right)-\eta_1^2 \left(5 (9+4 \nu
   )+4 \eta_2^2 (43+20 \nu )-10 \eta_2^4 (1-2 \nu
   )\right)\Biggr] \\ \notag
   f_V^{\rm inst,N^2LO}=&\frac{1}{420 \nu \left(1-\eta_1 \eta_2\right)^{2}}\Biggl\{7 \eta_2^2 \left(450 \nu -123+\eta_2^2 (70-80 \nu )+40
   \nu ^2+10 \eta_2^4 (-1+2 \nu )\right)\\ \notag
   &+70 \eta_1^6
   \left(2 \nu -1+\eta_2^2 (4-14 \nu )+\eta_2^4
   (8-27 \nu +4 \nu ^2)\right)\\ \notag
   &+\eta_1 \eta_2
   \Bigl(14 \eta_2^2 (189-700 \nu -80 \nu
   ^2)-1059-10920 \nu -560 \nu ^2+70 \eta_2^4 (-7+18 \nu
   )\Bigr)\\ \notag
   &+2 \eta_1^3 \Bigl(14 \eta_2^5 (23+70 \nu
   -40 \nu ^2)+7 \eta_2 (189-700 \nu -80 \nu
   ^2)+4 \eta_2^3 (184-5425 \nu -420 \nu
   ^2)\Bigr)\\ \notag
   &-\eta_1^5 \left(70 \eta_2 (7-18 \nu
   )-28 \eta_2^3 (23+70 \nu -40 \nu ^2)+\eta_2^5 (541+840 \nu +560 \nu ^2)\right)\\ \notag
   &+\eta_1^4
   \left(70 (7-8 \nu )+70 \eta_2^6 (8-27 \nu +4 \nu
   ^2)-7 \eta_2^2 (497-920 \nu -240 \nu ^2)-4
   \eta_2^4 (201-4130 \nu -560 \nu
   ^2)\right)\\ \notag
   &+\eta_1^2 \left(140 \eta_2^6 (2-7 \nu
   )-7 (123-450 \nu -40 \nu ^2)-7 \eta_2^4
   (497-920 \nu -240 \nu ^2)+4 \eta_2^2 (373+9800
   \nu +560 \nu ^2)\right)\\ \notag
   &+7 X_{12} \left(\eta_1^2-\eta_2^2\right) \Bigl(387-10 \eta_2^4+100 \nu
   +10 \eta_2^2 (7+8 \nu )+10 \eta_1^4 \left(\eta_2^2 (4+8 \nu )-1-2 \eta_2^4 (1-5 \nu )\right)\\ \notag
   &-\eta_1 \eta_2 \left(1377+560 \nu +10 \eta_2^2 (7+16 \nu
   )\right)-\eta_1^3 \eta_2 \left(70+160 \nu
   +\eta_2^2 (503+560 \nu )\right)\\ \label{eq: Vector_flux_explicit_expression_N${}^2$LO_coefficient}
   &+\eta_1^2 \bigl(70+80
   \nu +40 \eta_2^4 (1+2 \nu )+\eta_2^2 (1453+920 \nu
   )\bigr)\Bigr) \Biggr\}.
\end{align}
\end{subequations}

While the PN-coefficients of the tensor energy flux in Eqs.~\eqref{eq: tensor_flux_instantaneous_part} are
\begin{subequations}
\begin{align}
\notag
f_T^{\rm inst, NLO}=&\,- \frac{1}{336}\Biggl[1247+224 \eta_2^2+224 X_{12} \Bigl(\eta_1^2-\eta_2^2\Bigr)+980 \nu -2 \eta_1 \eta_2 (911+980 \nu
   )\\  \label{eq: Tensor_energy_flux_N${}^2$LO_coefficient}
   &\qquad +\eta_1^2 \Bigl(224+\eta_2^2 (127+980 \nu )\Bigr)
   \Biggr] \\  \notag
f_T^{\rm inst,N^2LO}=&\, -\frac{\left(1-\eta_1 \eta_2\right)^{-4/3}}{9072}\Biggl[44711+1008 \eta_2^4 (1-2 \nu )-18 \nu  (9271+1820 \nu ) +9 \eta_2^2 (-1053+5236 \nu )\\ \notag
&-9 X_{12} (\eta_1-\eta_2)
   (\eta_1+\eta_2) \Bigl(1053-112 \eta_1^2-426 \eta_1 \eta_2
   -112 \eta_2^2-403 \eta_1^2
   \eta_2^2+8876 (-1+\eta_1 \eta_2)^2 \nu \Bigr) \\ \notag
   &+\eta_1 \eta_2 \Bigl(-192803+18 \eta_2^2 (213-6244 \nu
   )+18 \nu  (41011+12860 \nu )\Bigr) \\ \notag
   &+\eta_1^4 \Bigl(1008 (1-2 \nu )+9 \eta_2^2 (403+7252 \nu )-4 \eta_2^4 \bigl(9196+9 \nu 
   (-2879+3700 \nu )\bigr)\Bigr)\\ \notag
   &+\eta_1^3 \eta_2 \Bigl(3834-112392 \nu +\eta_2^2 \bigl(-56021+18 \nu  (8937+24020 \nu )\bigr)\Bigr) \\ \label{eq: Tensor_energy_flux_NNLO_coefficient}
   &+3
   \eta_1^2 \Bigl(-3159+15708 \nu +3 \eta_2^4 (403+7252 \nu )+\eta_2^2 \bigl(80971-6 \nu  (46211+27660 \nu )\bigr)\Bigr)\Biggr].
\end{align} 
\end{subequations}

\section{Waveform modes }
\label{app: waveform_modes}
In Sec.~\ref{sec:waveformModes}, we derived the gravitational-wave emission from binaries of electrically charged black holes and, in particular, we presented the explicit expression for the harmonic mode $h_{22}$. In this appendix, we provide the explicit expressions for all relevant waveform harmonic modes up to 2PN order.

\subsection{Modes $\ell =2$}
\label{eq: l_2}
In the following, we present the $(2,1)$ mode as
\begin{equation}
    h^{21}= \frac{8}{3}\sqrt{\frac{\pi}{5}}\, i\, e^{-i\phi}\,M\, \nu\,x^{3/2}\Bigl\{X_{12}\left( 1-\eta_1 \eta_2\right)\,+x H^{21}_{\rm inst, NLO}\, +x^{3/2}\left(H^{21}_{\rm tail, LO}+ H^{21}_{\rm mem, LO}\right) \Bigr\}
\end{equation}
where
\begin{subequations}
   \begin{align}
   \notag
       &H^{21}_{\rm inst,NLO}=\,-\frac{x}{28\left(1-\eta_1 \eta_2\right)^{1/3}}\Bigl[14(1-4\nu)\left(\eta_1^2-\eta_2^2\right)\, \\
       & \hspace{2cm} +X_{12}\,\Bigl(17+14 \eta_2^2+14 \eta_1^2+10 \eta_2^2\eta_1^2 (1-2 \nu )-20 \nu -5 \eta_1 \eta_2 (11-8 \nu )\Bigr)\Bigr], \\ 
   &H^{21}_{\rm tail, LO}=\, \frac{1}{2} i x^{3/2} X_{12} (-1+\eta_1 \eta_2)
   \left(1+2 i \pi +4 \log (2)-6 \log \left(\frac{x}{x_0}\right)\right),\\
   &H^{21}_{\rm mem, LO}=\, \frac{i\, x^{3/2}}{48\left(1-\eta_1 \eta_2\right)} \Bigl[ \nu \left( \eta_1^2-\eta_2^2\right)\, -X_{12}\left(2 \eta_1 \eta_2(4-\nu)\, +\nu(\eta_1^2+\eta_2^2)\right)\Bigr].
   \end{align} 
\end{subequations}

\subsection{Modes $\ell =3$}
\label{eq: l_3}
In the following, we present all non-vanishing $\ell=3$ modes. 
\begin{equation}
    h^{33}=- i \sqrt{\frac{27\pi}{14}}\, M \nu\, e^{-3i\phi} \left(1-\eta_1\eta_2\right)\, x^{3/2}\Bigl\{ 2\, X_{12}\,+ x H^{33}_{\rm inst, NLO}\, +x^{3/2} \left(H^{33}_{\rm tail, LO}+ H^{33}_{\rm mem, LO}\right)\Bigr\}
\end{equation}
where
\begin{subequations}
    \begin{align}
        &H^{33}_{\rm inst,NLO}=\, \left(1-\eta_1 \eta_2\right)^{-4/3} \Bigl[X_{12}\left(\eta_1 \eta_2(11-6\nu)-\eta_1^2 -\eta_2^2-\eta_1^2 \eta_2^2 (1-2\nu)-8+4\nu\right)-(1-4\nu)(\eta_1^2-\eta_2^2)\Bigr] \\ 
        &H^{33}_{\rm tail, LO}=\, -\frac{6}{5}X_{12}\left[7i-5\pi+10i \log \left( \frac{3}{2}\right)+15i \log \left( \frac{x}{x_0}\right)\right] \\
        &H^{33}_{\rm mem, LO}=\, -\frac{i}{135}\Bigl[ 146\, \nu\,(\eta_1^2-\eta_2^2) -X_{12}\bigl(146 \nu (\eta_1^2+\eta_2^2)+\eta_1 \eta_2 (189-292\nu) \bigr)\Bigr]
    \end{align}
\end{subequations}

\begin{equation}
    h^{31}=\frac{1}{9}  \sqrt{\frac{\pi}{70}}\,i M \nu\, e^{-i\phi} \left(1-\eta_1\eta_2\right)\, x^{3/2}\Bigl\{ 6\, X_{12}\,+ x H^{31}_{\rm inst,NLO}\, +x^{3/2} \left(H^{31}_{\rm tail,LO}+ H^{33}_{\rm mem, LO}\right)\Bigr\}
\end{equation}
where
\begin{subequations}
    \begin{align}
        &H^{31}_{\rm inst, NLO}=\, -\frac{1}{\left(1-\eta_1 \eta_2\right)^{4/3}} \Bigl[3 (\eta_1^2-\eta_2^2)(1-4\nu)\,+X_{12}\bigl(16+4\nu+ 3\eta_1^2+3\eta_2^2-\eta_1 \eta_2(17+14\nu)-5 \eta_1^2 \eta_2^2(1-2\nu) \Bigr] \\ 
        &H^{31}_{\rm tail, LO}=\, -\frac{6}{5}X_{12}\left[7i-5\pi-i \log \left( 1024\right)+15i \log \left( \frac{x}{x_0}\right)\right] \\
        &H^{31}_{\rm mem, LO}=\, -\frac{i}{5}\Bigl[62 \nu (\eta_1^2-\eta_2^2)- X_{12}\bigl(62 \nu (\eta_1^2+\eta_2^2+\eta_1 \eta_2(7-124\nu) \bigr)\Bigr]
    \end{align}
\end{subequations}

\begin{equation}
    h^{32}=\frac{4}{135}  \sqrt{\frac{\pi}{7}}\, M \nu\, e^{-2i\phi}\, x^{2}\Bigl\{ H^{32}_{\rm inst, LO}\,+ x\, H^{32}_{\rm inst, NLO}\Bigr\}
\end{equation}
where
\begin{subequations}
    \begin{align}
        &H^{32}_{\rm inst,LO}=\,  90\, (1-\eta_1 \eta_2)^{4/3}(1-3\nu) \\ \notag
        &H^{32}_{\rm inst,NLO}=\, 193-725 \nu +365 \nu ^2+60 \eta_2^2 (1-3 \nu )+60 X_{12}
   \left(\eta_1^2-\eta_2^2\right) (1-3 \nu )\\ 
   &-2 \eta_1\eta_2 \left(193-725 \nu +365 \nu ^2\right)+60 \eta_1^2 (1-3 \nu )+73 \eta_1^2 \eta_2^2 \left(1-5 \nu +5
   \nu ^2\right)
    \end{align}
\end{subequations}

\subsection{Modes $\ell =4$}
\label{eq: l_4}
All non-vanishing $\ell=4$ modes are purely instantaneous and are provided only at LO, as follows
\begin{align}
    &h^{44}=\, -\frac{64}{9} \sqrt{\frac{\pi}{7}} M \nu\, x^2\, e^{-4i\phi}(1-\eta_1 \eta_2)^{4/3}(1-3\nu), \\
    &h^{43}=\, -\frac{9}{5} \sqrt{\frac{2\pi}{7}}\,i M \nu\, X_{12}\, x^{5/2}\, e^{-3i\phi}(1-\eta_1 \eta_2)^{5/3}(1-2\nu), \\
    &h^{42}=\, \frac{8}{63} \sqrt{\pi} M \nu\, x^2\, e^{-2i\phi}(1-\eta_1 \eta_2)^{4/3}(1-3\nu), \\
    &h^{41}=\, \frac{1}{105} \sqrt{2 \pi}\, i M \nu\, X_{12}\, x^{5/2}\, e^{-i\phi}(1-\eta_1 \eta_2)^{5/3}(1-2\nu).
\end{align}

\subsection{Modes $\ell =5$}
\label{eq: l_5}
All non-vanishing $\ell=5$ modes are purely instantaneous and are provided only at LO, as follows
\begin{align}
    &h^{55}=\, \frac{125}{12} \sqrt{\frac{5\pi}{56}}\, i M \nu\, X_{12}\, x^{5/2}\, e^{-5i\phi}(1-\eta_1 \eta_2)^{5/3}(1-2\nu), \\
    &h^{54}=\, -\frac{256}{45} \sqrt{\frac{\pi}{33}} M \nu\, x^2\, e^{-4i\phi}(1-\eta_1 \eta_2)^{2}(1-5\nu+5\nu^2), \\
    &h^{53}=\, -\frac{9}{20} \sqrt{\frac{3\pi}{22}}\,i M \nu\, X_{12}\, x^{5/2}\, e^{-3i\phi}(1-\eta_1 \eta_2)^{5/3}(1-2\nu), \\
    &h^{52}=\, \frac{16}{135} \sqrt{\frac{\pi}{11}}\, M \nu\, x^3\, e^{-2i\phi}(1-\eta_1 \eta_2)^{2}(1-5\nu+5\nu^2), \\
    &h^{51}=\, \frac{1}{180} \sqrt{\frac{\pi}{77}}\, i M \nu\, X_{12}\, x^{5/2}\, e^{-i\phi}(1-\eta_1 \eta_2)^{5/3}(1-2\nu).
\end{align}

\subsection{Modes $\ell =6$}
\label{eq: l_6}
All non-vanishing $\ell=6$ modes are purely instantaneous and are provided only at LO, as follows
\begin{align}
    &h^{66}=\, \frac{434}{5} \sqrt{\frac{\pi}{715}}\, M \nu\, x^3\, e^{-6i\phi}(1-\eta_1 \eta_2)^{2}(1-5\nu+5\nu^2), \\
    &h^{64}=\, -\frac{1024}{495} \sqrt{\frac{2\pi}{195}}\, M \nu\, x^3\, e^{-4i\phi}(1-\eta_1 \eta_2)^{2}(1-5\nu+5\nu^2), \\ 
    &h^{62}=\, \frac{16}{1485} \sqrt{\frac{\pi}{13}}\, M \nu\, x^3\, e^{-2i\phi}(1-\eta_1 \eta_2)^{2}(1-5\nu+5\nu^2).
\end{align}
\end{widetext}

\newpage
\bibliography{bibliography.bib} 

\end{document}